\documentclass[a4paper]{article}

\usepackage{siunitx}
\usepackage{subfigure}
\usepackage{bm}
\usepackage{moreverb}
\usepackage{diagbox}
\usepackage{multirow}
\usepackage{authblk}
\usepackage{amsmath}
\usepackage{pdfpages}
\usepackage{geometry}

\usepackage[driverfallback=dvipdfm,colorlinks,bookmarksopen,bookmarksnumbered,citecolor=red,urlcolor=red]{hyperref}

\usepackage{tikz}
\usetikzlibrary{
  calc,%
  trees,%
  positioning,%
  arrows,%
  chains,%
  shapes.geometric,%
  decorations.pathreplacing,%
 decorations.pathmorphing,%
  shapes,%
  matrix,%
 shapes.symbols,%
 intersections,%
  fit,%
  patterns,%
  angles,%
  quotes%
 }
\usepackage{pgfplots}
\usepackage{pgfkeys}
\pgfplotsset{compat=newest}
\pgfkeys{/pgf/number format/1000 sep=}
\usepackage{pgfplotstable}
\pgfplotsset{
/pgfplots/colormap={gray}{rgb255=(0,0,0) rgb255=(255,255,255)}
}
\pgfplotsset{
/pgfplots/colormap={grayinv}{rgb255=(255,255,255) rgb255=(0,0,0)}
}

\tikzstyle{startstop} = [rectangle, rounded corners, minimum width=3cm, minimum height=1cm,text centered, draw=black]
\tikzstyle{io} = [trapezium, trapezium left angle=70, trapezium right angle=110, minimum width=3cm, minimum height=1cm, text centered, draw=black]
\tikzstyle{arrow} = [thick,->,>=stealth]
\tikzstyle{decision} = [diamond, draw,minimum width=2.5cm,minimum height=1.5cm text width=4.5em, text centered, inner sep=0pt]

\definecolor{color1}{HTML}{0000AD} 
\definecolor{color2}{HTML}{FF4500} 
\definecolor{color3}{HTML}{FFA500} 
\definecolor{color4}{HTML}{00BB00} 
\definecolor{color5}{HTML}{9400D3} 
\definecolor{color6}{HTML}{800000} 
\definecolor{color7}{HTML}{000000} 
\definecolor{color8}{HTML}{0000FF} 
\definecolor{color9}{HTML}{FF0000} 
\definecolor{color10}{HTML}{11CCEE} 
\definecolor{color11}{HTML}{606060} 

\tikzset{%
    body/.style={inner sep=0pt,outer sep=0pt,shape=rectangle,draw,thick,pattern=north east lines wide},
    dimen/.style={<->,>=latex,thin,every rectangle node/.style={fill=white,midway}},
    symmetry/.style={dashed,thin},
}

\tikzstyle{line1} = [color=color7,semithick]
\tikzstyle{line2} = [color=color2,densely dotted,thick]
\tikzstyle{line3} = [color=color1,densely dashed,thick]
\tikzstyle{line4} = [color=color4,dash dot,thick]
\tikzstyle{line5} = [color=color5,dash dot dot,thick]
\tikzstyle{line6} = [color=color6,densely dotted,thick]
\tikzstyle{line7} = [color=color10,densely dashed,thick]
\tikzstyle{line8} = [color=color3,thick]

\tikzstyle{mark1} = [color=color7,mark=x,mark size=2pt,mark options=solid,semithick]
\tikzstyle{mark2} = [color=color2,mark=square,mark size=2pt,mark options=solid,semithick]
\tikzstyle{mark3} = [color=color1,mark=triangle,mark size=2pt,mark options=solid,semithick]
\tikzstyle{mark4} = [color=color4,mark=o,mark size=2pt,mark options=solid,semithick]
\tikzstyle{mark5} = [color=color2,mark=square*,mark size=2pt,mark options=solid,semithick]
\tikzstyle{mark6} = [color=color1,mark=triangle*,mark size=2pt,mark options=solid,semithick]
\tikzstyle{mark7} = [color=color4,mark=*,mark size=2pt,mark options=solid,semithick]
\tikzstyle{mark8} = [color=color2,mark=x,mark size=2.5pt,mark options=solid,semithick]
\tikzstyle{mark9} = [color=color1,mark=diamond,mark size=2pt,mark options=solid,semithick]
\tikzstyle{mark10} = [color=color4,mark=asterisk,mark size=2.5pt,mark options=solid,semithick]
\tikzstyle{mark11} = [color=color7,mark=*,mark size=2pt,mark options=solid,semithick]
\tikzstyle{mark12} = [color=color5,mark=*,mark size=2pt,mark options=solid,semithick]
\tikzstyle{mark13} = [color=color7,mark=*,mark size=1pt,mark options=solid,semithick]
\tikzstyle{mark14} = [color=color3,mark=x,mark size=2.5pt,mark options=solid,semithick]

\pgfplotsset{major grid style={densely dotted}} 

\newcommand\BibTeX{{\rmfamily B\kern-.05em \textsc{i\kern-.025em b}\kern-.08em
T\kern-.1667em\lower.7ex\hbox{E}\kern-.125emX}}

\begin{document}

\title{Free surface flow through rigid porous media - An overview and comparison of formulations}
\author[1]{W. D\"usterh\"oft-Wriggers}
\author[2]{A. Larese}
\author[3 , 4]{E. O\~{n}ate}
\author[1]{T. Rung}
\affil[1]{Hamburg University of Technology, Hamburg, Germany}
\affil[2]{Universit\`a degli Studi di Padova, Padova, Italy}
\affil[3]{International Center for Numerical Methods in Engineering, CIMNE}
\affil[4]{Universitat Polit\'ecnica de Catalunya, Barcelona, Spain}
\affil[]{corresponding author: wibke.wriggers@tuhh.de}

\date{}

\renewcommand\Affilfont{\itshape\small}
\maketitle

\begin{abstract}In many applications free surface flow through rigid porous media has to be modeled. Examples refer to coastal engineering applications as well as geotechnical or biomedical applications. Albeit the frequent applications, slight inconsistencies in the formulation of the governing equations can be found in the literature. The main goal of this paper is to identify these differences and provide a quantitative assessment of different approaches. 
Following a review of the different formulations, simulation results obtained from three alternative formulations are compared with experimental and numerical data. Results 
obtained by 2D and 3D test cases indicate that the predictive differences returned by the different formulations remain small for most applications, in particular for small porous Reynolds number $Re_P< 5000$. Thus it seems justified to select a formulation that supports an efficient algorithm and coding structure. 
\end{abstract}


\vspace{-6pt}

\section{Introduction}
\vspace{-2pt}
Geotechnical and coastal engineering applications often require simulation procedures that can model partially and fully saturated granular media. Also in e.g. biomedical engineering and in  manufacturing, the flow through porous materials has to be considered. In this paper the governing equations for the simulation of incompressible, immiscible two-phase flow through rigid porous media are explored. Studying the literature, it is found that different formulations exist, e.g. \cite{DeLemos2006}, \cite{DelJesus2012}, \cite{Hsu2002}, \cite{Larese2015},  \cite{Liu1999}, \cite{Uzuoka2012}, \cite{VanGent1992}. The differences primarily refer to the transport terms of the momentum equation and are particularly influential at the boundary of the porous material or for an inhomogeneous porosity.  To assess the predictive discrepancies in formerly investigated 2D and 3D test cases, a unified formulation is sought in this work.\\
Parts of the deviations addressed herein were already outlined by del Jesus et al \cite{DelJesus2012}. The related discussion of equations will be extended in this work.
The approach of this paper is to firstly list variations of the governing equations from the literature. Subsequently, two options are implemented in a cell-centered Finite Volume solver using a volume of fluid approach to solve the two-phase flow. Moreover, results of a third edge-based formulation by the authors \cite{Larese2015} are compared. 
The three formulations are applied to illustrative 2D and 3D cases, which feature  different porous Reynolds numbers and results are discussed. Results reveal, that certain simplifications can be introduced without a traceable predicitve quality change.\\
The paper aims to contribute to the understanding of dominant influences in the equations for a two-phase flow through rigid porous medium. Furthermore, the comparison of test case results for different methods and formulations can help establishing benchmark cases for such problems.
The paper is structured as follows: In section \ref{ThBackgr} the governing equations are discussed in the context of the existing literature. Different notations are compared and two different equation sets to be tested are derived from these considerations. 
Section \ref{NUM} outlines the numerical treatment for a cell-centered, second-order accurate Finite Volume method. Applications of the different formulations are discussed in section \ref{2DTestcases}, which also ranks the formulation related influences against resolution, approximation and porous resistance force modelling aspects. Final conclusions are drawn in section \ref{Concl}.\\
The notation uses standard lower case Latin subscripts to mark Cartesian tensor
coordinates and Einsteins summation is used over repeated lower case indices. Symbolic notations of vectors and tensors employ underlining (e.g. $\underline v$). 

\vspace{-6pt}

\section{Mathematical Model}
\label{ThBackgr}
\begin{figure}[htbp]
\begin{center}
\subfigure[Variable $v^F_i$ and $p^F$]{ 
 \includegraphics{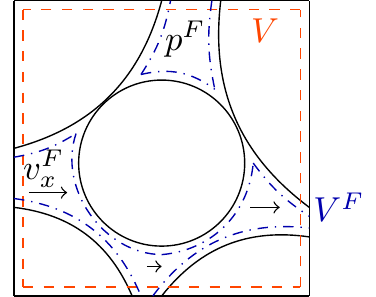}
  }
  \subfigure[Variable $\hat{v}_i$ and $p^F$]{ 
 \includegraphics{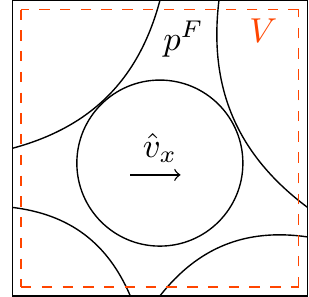}
   }
  \end{center}
  \caption{Schematic view of flow through porous medium.}
   \label{CellVelocityPressureFluidBild}
\end{figure}

\vspace{-2pt}
The simulation of two-phase flow through rigid porous material is simulated solving the Navier-Stokes equations in an Eulerian framework, modified so to account for the presence of the porous media. The present paper studies only immiscible fluids featuring incompressible bulk fluids and a sharp interface. In such cases, Volume of Fluid  (VoF) or level set (e.g. \cite{Hirt1981}, \cite{Sussmann1994}) interface capturing approaches, which assume a unique velocity field, are  frequently used simpler alternatives. They  involve one continuity and one momentum equation for the mixture, which are supplemented by an equation to identify the local fluid phase and simple equations of state for the mixture properties. Hence, VoF and level set results are the basis of displayed computed own results.
\\
Following the literature, different formulations of the governing equations are discussed in this section. We start with the definition of a porosity $n$, defined as the ratio of the fluid volume $V^F$ to the total volume $V$
\begin{equation}
n=\frac{V^F}{V}\;.
\end{equation}
The total volume $dV$ consists of the fluid volume $dV^F$ and the volume of the rigid porous material $dV^S$ ($dV=dV^F+dV^S$), cf.  
Figure \ref{CellVelocityPressureFluidBild}.\\
The flow through a porous medium can either be expressed in terms of fluid velocity $v^F_i$ or of the Darcy velocity $\hat{v}_i$. 
The Darcy velocity can be obtained from the fluid velocity by multiplication with the porosity $n$,  i.e. $\hat{v}_i=n \, v^F_i$.

\subsection{Continuity equation}
\label{ConEq}
Considering the mass conservation of the fluid inside the fluid volume for an incompressible fluid   
\begin{equation}
\frac{D}{Dt}\int_{V^F} \rho dV=\int_{V^F} \left(\frac{\partial \rho}{\partial t} +\frac{\partial v^F_i \rho}{\partial x_i}\right) dV=0
\label{MassErhVol}
\end{equation}
leads to the usual zero-divergence differential form for one-phase flows with constant density $\rho$
\begin{equation}
\int_{V^F}\frac{\partial v^F_i}{\partial x_i} dV=0\;.
\label{FluidVolConti}
\end{equation}
Introducing immiscible two-phase flow (i.e. air and water) through a rigid porous medium and using a VoF approach, the density is obtained from an air mixture fraction $c=V^{air,F}$/$V^F$
\begin{equation}
\rho=c \rho^A+(1-c) \rho^W
\label{stoffgesetzDens}
\end{equation}
with constant bulk densities $\rho^A$ and $\rho^W$. Introducing \eqref{stoffgesetzDens} in \eqref{MassErhVol}, again the zero-divergence differential form \eqref{FluidVolConti} follows when the immiscible condition $D c$/$D t=0$ is applied.\\
Integrating over the total volume $V$ instead of the fluid volume $V^F=n V$ and introducing the Darcy velocity $\hat{v}_i=n v^F_i$, equation \eqref{FluidVolConti} gets   
\begin{equation}
\int_{V} \left(n \frac{\partial \frac{\hat{v}_i}{n}}{\partial x_i}\right) dV =\oint_{A} \hat{v}_i dA_i-\int_{V} \left( \frac{\partial n}{\partial x_i}\frac{\hat{v}_i}{n}\right) dV=0
 \label{contiVolV}
\end{equation}
as is also stated in \cite{DelJesus2012}. The second term obviously vanishes for spatially constant porosities. If the simulation domain includes a porous material with constant porosity and an area without porous material, the second term only influences the results at the boundary of the porous material. To study the influence of the momentum equation formulations, the continuity equation is applied as in, e.g. Larese et. al \cite{Larese2015} or Hsu et. al \cite{Hsu2002} where the second term is neglected and therefore the divergence of the Darcy velocity $\hat{v}_i$ vanishes
\begin{equation}
\int_{V}\frac{\partial \; \hat{v}_i }{\partial x_i}dV=0\;.
\label{WholeVolConti}
\end{equation}

\subsection{Momentum equation}
\label{MomEQTHEO}
As a starting point, the well known momentum equation of a fluid inside a fluid volume 
is used to derive the momentum equation in porous media.
For each fluid volume $d V^F$ 
\begin{equation}
\left(\frac{\partial (\rho v_i^F)}{\partial t}+\frac{\partial \left(\rho v_i^F v_j^F\right)}{\partial x_j}\right) dV^F=\rho g_i dV^F- \frac{\partial p^F}{\partial x_i}dV^F+\frac{\partial }{\partial x_i}\left(\mu \frac{\partial v_i^F}{\partial x_j} \right) dV^F\;.
\label{MomTheorCons}
\end{equation}
Formulating the momentum equation for an incompressible fluid over the complete control volume $dV=dV^F$/$n$, introducing the Darcy velocity $\hat{v}_i$ as well as porous resistance forces $\hat{f}^R$ and dividing by the constant density leads to
\begin{equation}
\underbrace{n\frac{\partial (v_i^F)}{\partial t}}_{a^I_i}+\underbrace{ \hat{v}_j \frac{\partial \left( v_i^F \right)}{\partial x_j}}_{a^C_i} = \underbrace{n g_i}_{\hat{f}^G}  \underbrace{- \frac{n}{\rho}\frac{\partial p^F}{\partial x_i}}_{\hat{f}^P} +\underbrace{\frac{n}{\rho}\frac{\partial }{\partial x_i}\left(\mu \frac{\partial v_i^F}{\partial x_j} \right) }_{-a^D_i}+\hat{f}^R_i \;.
\label{MomTheor}
\end{equation}
The porous momentum equations are stated with minor modifications in the existing literature. An overview of the respective terms of a generic momentum equation (equation \eqref{MomTheor}) is given in Tables \ref{MomEqVerglLS} and \ref{MomEqVerglRS} for a selection of publications. 
In order to compare the
differences exposed by these formulations, all equations are written in a unified form.
The selected generic momentum equation refers to equation \eqref{MomTheor} which consists of the inertia term of the Darcy velocity $a^I_i$ in addition to a convective $a^C_i$ and a diffusion term $a^D_i$. These terms are balanced by a pressure force $\hat{f}^P_i$, a gravity force $\hat{f}^G_i$ and a porous resistance $\hat{f}^R_i$.

\begin{table}[h]
\caption{Literature reported left hand side terms of the momentum equation}
\begin{center}
\begin{tabular}{|c|c|c|c|c|c|c|}
\hline
literature&$\bm{a^I_i}$&$\bm{a^C_i}$&$\bm{a^D_i}$\\
\hline
Uzuoka et. al \cite{Uzuoka2012}&$n \frac{\partial v_i^F}{\partial t}$&\multirow{3}{*}{$\hat{v}_j \frac{\partial}{\partial x_j}\left(v_i^F\right)$}&-\\
\cline{1-2} \cline{4-4}
del Jesus \cite{DelJesus2012}&\multirow{6}{*}{$\frac{\partial \hat{v_i}}{\partial t}$}&&$-\frac{n}{\rho}\frac{\partial}{\partial x_i}\left(\mu \frac{\partial}{\partial x_j}v^F_i\right)$\\
\cline{1-1} \cline{4-4}
De Lemos \cite{DeLemos2006}&&&$-\frac{1}{\rho}\frac{\partial}{\partial x_i}\left(\mu \frac{\partial}{\partial x_j}\hat{v}_i\right)$\\ 
\cline{1-1} \cline{3-4}
Larese et. al \cite{Larese2015}&&\multirow{4}{*}{$v_j^F \frac{\partial}{\partial x_j}\left(\hat{v}_i\right)$}&$-\frac{\partial}{\partial x_i}\left(\frac{\mu}{\rho} \frac{\partial}{\partial x_j} \hat{v}_i\right)$\\
\cline{1-1} \cline{4-4}
Hsu et. al \cite{Hsu2002}&&&$-\frac{1}{\rho}\frac{\partial}{\partial x_i}\left(\mu \frac{\partial}{\partial x_j}\hat{v}_i\right)$\\
\cline{1-1} \cline{4-4}
Liu et. al \cite{Liu1999}&&&$-\frac{\mu}{\rho}\frac{\partial^2}{\partial x_j^2}\left(\hat{v}_i\right)$\\
\cline{1-1} \cline{4-4}
Van Gent \cite{VanGent1992} (1D)&&&-\\
\hline
\end{tabular}
\end{center}
\label{MomEqVerglLS}
\end{table}

\begin{table}[h]
\caption{Literature reported right hand side terms of the momentum equation}
\begin{center}
\begin{tabular}{|c|c|c|c|c|c|c|}
\hline
literature&$\bm{\hat{f}^P_i}$&$\bm{\hat{f}^G_i}$&$\bm{\hat{f}^R_i}$\\
\hline
Uzuoka et. al \cite{Uzuoka2012}&\multirow{2}{*}{$-\frac{n}{\rho}\frac{\partial \left( p^F\right)}{\partial x_i}$}&\multirow{5}{*}{$n g_i$}&$-\tilde A \hat{v}_i$\\
\cline{1-1} \cline{4-4}
del Jesus \cite{DelJesus2012}&&&$-\tilde A \hat{v}_i -\tilde B \hat{v}_i \mid \underline{\hat{v}} \mid- C\frac{\partial}{\partial t}\hat{v}_i$\\
\cline{1-2} \cline{4-4}
De Lemos \cite{DeLemos2006}&$-\frac{1}{\rho}\frac{\partial}{\partial x_i}\left(n p^F\right)$&&\multirow{2}{*}{$-\tilde A \hat{v}_i-\tilde B \hat{v}_i \mid \underline{\hat{v}} \mid$}\\
\cline{1-2} 
Larese et. al \cite{Larese2015}&\multirow{4}{*}{$-\frac{n}{\rho}\frac{\partial \left( p^F\right)}{\partial x_i}$}&&\\
\cline{1-1} \cline{4-4}
Hsu et. al \cite{Hsu2002}&&&\multirow{3}{*}{$-\tilde A \hat{v}_i- \tilde B \hat{v}_i \mid \underline{\hat{v}} \mid- C\frac{\partial}{\partial t}\hat{v}_i$}\\
\cline{1-1} \cline{3-3}
Liu et. al \cite{Liu1999}&&-&\\
\cline{1-1} \cline{3-3}
Van Gent \cite{VanGent1992} (1D)&&$n g_i $&\\
\hline
\end{tabular}
\end{center}
\label{MomEqVerglRS}
\end{table}
Note that all the authors mentioned in Table \ref{MomEqVerglLS} and \ref{MomEqVerglRS} had different applications in mind. Hsu et. al \cite{Hsu2002}, Liu et. al \cite{Liu1999} and Del Jesus \cite{DelJesus2012} address the numerical simulations of wave motions close to and through porous coastal structures. Also the work of Larese et. al \cite{Larese2015} is about water flow through rockfill and rockfill-like materials. Van Gent \cite{VanGent1992} focusses on theoretical aspects of formulas to describe porous flows and applicable resistance laws. The equations are explored in one dimension. Uzuoka et. al \cite{Uzuoka2012} have a completely different target since they concentrate on the deformation of unsaturated poroelastic solids. They use the momentum equation above in their derivation of their own Lagrangian approach to simulate poroelastic solids. De Lemos \cite{DeLemos2006} writes generally about turbulent flow in saturated rigid porous media. These different backgrounds do partially explain the listed variations in the governing equations.\\
Three sets of dependent variables ($v_i^F$, $p^F$), ($\hat{v}_i$, $p^F$) and ($\hat{v}_i$, $\hat{p}$) are employed in the mentioned studies. Uzuoka et. al \cite{Uzuoka2012} use the fluid velocity $v_i^F$ and fluid pressure $p^F$ as variables, Larese et. al \cite{Larese2015}, Liu et. al \cite{Liu1999}, De Lemos \cite{DeLemos2006}, Hsu et. al \cite{Hsu2002} and Van Gent \cite{VanGent1992} are deploying the Darcy velocity $\hat{v}_i$ and fluid pressure $p^F$ and Del Jesus \cite{DelJesus2012} refers to the Darcy velocity $\hat{v}_i$ and the ensemble-averaged pressure $\hat{p}$  which can be translated in the fluid pressure by $\hat{p}=n p^F$.\\
The formulation of the inertia term of Uzuoka et. al \cite{Uzuoka2012} considers variable porosities whereas all other formulations are applied to rigid porosity materials with constant porosity.\\
Two different versions of the convective term exist, one  based on the divergence of the fluid velocity (Uzuoka et. al \cite{Uzuoka2012}, Del Jesus \cite{DelJesus2012}, De Lemos \cite{DeLemos2006}) and one using the divergence of the Darcy velocity (Larese et. al \cite{Larese2015}, Hsu et. al \cite{Hsu2002}, Liu et. al \cite{Liu1999}, Van Gent \cite{VanGent1992}). Both formulations only agree if the porosity is homogeneous. 
For a spatially non-constant porosity, the first version follows from the derivation of the porous momentum equations, see above or e.g. Del Jesus \cite{DelJesus2012}. The second version is derived by considering the fluid velocity transporting the Darcy momentum. This imposes a lack of momentum transport at the boundaries of the porous material as highlighted in \cite{DelJesus2012}. The amount of error depends on the face interpolation of the porosity in this case.
\\
The fluid friction within the porous material is neglected by Uzuoka et. al \cite{Uzuoka2012}. 
This is related to the focus of their simulations on predicting the deformation of poroelastic solids rather than the fluid flow inside the solid. Uzuoka et. al \cite{Uzuoka2012} concentrate on materials with a low intrinsic permeability. Van Gent \cite{VanGent1992} neglects the viscous term of the fluid flow and instead includes related influences in a non-linear term of the porous resistance. Except for minor differences as regards the fluid properties, all other formulations besides Del Jesus \cite{DelJesus2012} match in the diffusion term. Del Jesus et al.  \cite{DelJesus2012} consider the  porosity gradient in the diffusion term, whereas the other approaches are formulated for a spatially constant porosity. 
\\
The pressure term coincides for Uzuoka et. al \cite{Uzuoka2012}, Del Jesus \cite{DelJesus2012}, Hsu et. al \cite{Hsu2002}, Larese et. al \cite{Larese2015} and Van Gent \cite{VanGent1992}. In Del Jesus \cite{DelJesus2012} it is stated that Hsu et. al \cite{Hsu2002} use a different pressure term which gives a smaller pressure drop. This cannot be confirmed by the present results. The formulation of De Lemos \cite{DeLemos2006} again coincides with the other formulations for a spatially constant porosity.
\\
The tree main variants of the porous resistance force are used by the tabulated formulations:   
Darcy's formula (Uzuoka et. al \cite{Uzuoka2012}), a Forchheimer formula (Larese et. al \cite{Larese2015}, De Lemos \cite{DeLemos2006}) and a Forchheimer formula with an added mass term (Del Jesus \cite{DelJesus2012}, Hsu et. al \cite{Hsu2002}, Liu et. al \cite{Liu1999}, Van Gent \cite{VanGent1992}). The  use of the formulation depends on the nature of the investigated flow field which can be described by the pore Reynolds number. A detailed description of resistance laws and their range of application is given in \cite{VanGent1992}. Further details will be provided in section \ref{PorForces}. \\
In order to  compare different versions of the equation system, two momentum equations featuring different sets of variables are discretized in this work. Using the unified form of the equations, i.e.  
\begin{equation}
 \frac{\partial \hat{v}_i}{\partial t}+\hat{v}_j \frac{\partial}{\partial x_j}\left(\frac{1}{n}\hat{v}_i\right)-\frac{1}{\rho}\frac{\partial}{\partial x_i}\left(\mu \frac{\partial}{\partial x_j}\hat{v}_i\right)=-\frac{n}{\rho}\frac{\partial \, p^F }{\partial x_i}+n g_i- \tilde A \hat{v}_i- \tilde B \hat{v}_i \mid \underline{\hat{v}} \mid
 \label{XMomEq1a}
\end{equation}
and
\begin{equation}
 n \frac{\partial v^F_i}{\partial t}+ n v^F_j \frac{\partial}{\partial x_j}\left(v^F_i\right)-\frac{n}{\rho}\frac{\partial}{\partial x_i}\left(\mu \frac{\partial}{\partial x_j}v^F_i\right)=-\frac{n}{\rho}\frac{\partial \, p^F }{\partial x_i}+n g_i-n \tilde A v^F_i- n^2 \tilde B v^F_i \mid  \underline{{v}}^F \mid\;.
  \label{XMomEq2a}
\end{equation}
Apart from the usage of a different velocity formulation, the equations differ in the diffusion term where equation \eqref{XMomEq1a} includes the divergence of the porosity hidden inside the Darcy velocity. In the diffusion term in equation \eqref{XMomEq2a} the porosity is excluded and therefore the divergence of it is not considered. This difference is especially relevant for non-constant porosities as well as at the boundaries of porous materials. For temporally inconstant porosities also the time term of equation \eqref{XMomEq1a} and \eqref{XMomEq2a} leads to a different discretisation.\\
Additionally numerical results are compared against the results from \cite{Larese2015}, which provides a third version of the governing equations. Equation (\ref{XMomEq1a}) coincides with the formulation of Del Jesus \cite{DelJesus2012} in all terms except the diffusion term. Equation (\ref{XMomEq2a}) is an extension of the formulation of Uzuoka et. al \cite{Uzuoka2012} with a diffusion term that matches the one of Del Jesus \cite{DelJesus2012} in accord with the Forchheimer law. The equation system used by \cite{Larese2015} only differs from  (\ref{XMomEq1a}) in the convective term.

\subsection{Porous forces}
\label{PorForces}
The analysis of the flow through a porous media is well established in an engineering context. The initial model relates to Darcy's linear law, i.e. $\hat f_i^R \sim - \tilde A \hat v_i$ which is known since 1856. Its range of applicability is discussed in length by \cite{PolubarinovaKochina1952}, \cite{Larese2013} and \cite{VanGent1995}. For small porous Reynolds numbers
\begin{equation}
Re_P=\frac{\mid \underline{\hat{v}} \mid D_{50}}{\nu} \, ,
\end{equation}
where $D_{50}$ refers to the mean grain size of the porous material,  a linear law is sufficient to represent the resistance forces acting on the fluid. Non-linear behaviour starts at $Re_P \in [1-10]$, see \cite{GuWang1991}, and turbulence appears around $Re_P \in [60-150]$ as stated in \cite{Larese2013}. In 1901 Darcy's law was extended by Forchheimer \cite{Forchheimer1901} using a nonlinear term to represent the contribution of turbulence and parts of large-scale convective transport. An additional transient term was added by Polubarinova-Kochina in \cite{PolubarinovaKochina1952} which takes added mass phenomena into account. The latter leads to an extended Forchheimer equation, viz. 
\begin{equation}
\hat{f}^R_i=- \left( \tilde A \hat{v}_i + \tilde B \hat{v}_i \mid \underline{\hat{v}} \mid+ \tilde C\frac{\partial \hat{v}_i}{\partial t} \right) \; . 
\label{Forchheim}
\end{equation}
Expression \eqref{Forchheim} still represents a state-of-the-art approach for flows through porous material. The resistance force factors $\tilde A$, $\tilde B$ and $\tilde C$ are best obtained by experiments. Many authors tried to derive general expressions which depend on the porosity $n$, the physical properties of the fluid, the mean grain diameter $D_{50}$ and additional constant, cf. \cite{Bear1972}. Table \ref{PorForceVergl} refers to the models used by the previously mentioned numerical simulations.
\begin{table}[h]
\caption{Different formulations of resistance law constants used in the literature}
\begin{center}
\begin{tabular}{|c|c|c|c|c|}
\hline
literature&type of res. law&$\tilde A$&$\tilde B$&$\tilde C$\\
\hline
Uzuoka et. al \cite{Uzuoka2012}& general Darcy&\multirow{2}{*}{$\frac{\mu n}{\rho k}$}&-&-\\
\cline{1-2} \cline{4-5} 
De Lemos \cite{DeLemos2001}&general nonlinear&&$ \frac{c_F n}{\sqrt{k}}$&-\\
\hline
Larese et. al \cite{Larese2013} &Ergun&\multirow{2}{*}{$\alpha \frac{\mu \left(1-n\right)^2}{\rho n^2 D_{50}^2}$}&\multirow{4}{*}{$\beta \frac{ \left(1-n\right)}{n^2 D_{50}}$}&-\\
\& De Lemos \cite{DeLemos2006}&&&&\\
\cline{1-2} \cline{5-5} 
Van Gent \cite{VanGent1992} (1D) &Van Gent&&&\multirow{3}{*}{$\gamma \frac{\left(1-n\right)}{n}$}\\
\& Hsu et. al \cite{Hsu2002}&&&&\\
\cline{1-3} 
del Jesus \cite{DelJesus2012} \&  \cite{VanGent1992} &Engelund&$\alpha\frac{ \mu \left(1-n\right)^3}{\rho n D^2_{50}}$&&\\
\cline{1-4} 
Liu et. al \cite{Liu1999}&Van Gent&$\alpha \frac{\mu \left(1-n\right)^2}{\rho n^2 D_{50}^2}$&$\beta\frac{ \left(1+\frac{7.5}{KC}\right)\left(1-n\right) }{n^2 D_{50}}$&\\
&oscillating flow&&&\\
\hline
\end{tabular}
\end{center}
\label{PorForceVergl}
\end{table}
Uzuoka et. al \cite{Uzuoka2012} use Darcy's law in a general form where $k$ is the intrinsic permeability. In Uzuoka et. al \cite{Uzuoka2012} originially the hydraulic conductivity $K$ is used and obtained from a model which depends on the water saturation of the two-phase fluid. A clear relation between the hydraulic conductivity and the intrinsic permeability is given by $K=\frac{\rho}{\mu}g k$\;. Larese et al. \cite{Larese2013} and De Lemos et al. \cite{DeLemos2006}  employ an Ergun type of model. In \cite{DeLemos2001} De Lemos et al. refer to a general expression for $\tilde A$ and $\tilde B$ based upon the intrinsic permeability $k$ which can be obtained from experiments or one of the aforementioned models. 
The Van Gent model is used by Liu et al. \cite{Liu1999} and Van Gent et al. \cite{VanGent1992}, where the former employs a more complex version which adjusts $\tilde A$ and $\tilde B$ with the help of the Keuler-Carpenter number $KC=\frac{\mid \hat{v}_i T\mid}{D_{50}}$ an a typical time scale $T$. The Van Gent model is also used in \cite{Hsu2002}. Van Gent et al. \cite{VanGent1992} also used the Engelund model for verification purposes. With the Engelund formula given in table \ref{PorForceVergl} a recalculation of \cite{DelJesus2012} results are possible.  
 
\subsection{Two-phase model}
\label{FreeSurfEQ}
Table \ref{FreeSurfEqVergl} outlined the two-phase methods applied by the reference publications. As seen from the table, the majority of authors used an interface capturing approach \cite{Hirt1981, Sussmann1994, Anderson1998} which identifies the position of the interface from a
scalar indicator or concentration or phase field function, cf. also related reviews in \cite{Mirjalili2017} and \cite{Tryggvason2011}. Table \ref{FreeSurfEqVergl} also provides a brief background of the employed numerical methods.
\begin{table}[h]
\caption{Methods used for free surface flow in literature}
\begin{center}
\begin{tabular}{|c|c|c|}
\hline
literature&interface method&discretisation method\\
\hline
Uzuoka et. al \cite{Uzuoka2012}& Euler-Euler &Finite Elements\\
\hline
Larese et. al \cite{Larese2015}&level set&edge based fractional step solver\\
\hline
del Jesus \cite{DelJesus2012}&VOF&Finite Volume\\
\hline
Hsu et. al \cite{Hsu2002}&VOF&Finite Differences\\
\hline
De Lemos \cite{DeLemos2006}&black oil model&IMPES\\
\hline
Liu et. al \cite{Liu1999}&VOF&Finite Differences\\
\hline
Van Gent \cite{VanGent1992} &shallow water equations&Finite Differences\\
\hline
\end{tabular}
\end{center}
\label{FreeSurfEqVergl}
\end{table}
Larese et. al \cite{Larese2015} rely on the level set method to detect the phases and only solve the momentum and pressure equation for the water phase. The approach offers the advantage of an inherently sharp interface for immiscible flows.  Van Gent \cite{VanGent1992} also restricts the simulation to the water phase by applying shallow water equations. In contrast to all other authors, Uzuoka et. al \cite{Uzuoka2012} solve governing equations for both fluid phases. In this work, the fluid is assumed to consist of two immiscible phases, i.e. air and water. Using a VoF approach, the fluid properties are obtained from an air mixture fraction $c=V^{air,F}/V^F$ which determines the air occupied volume $V^{air,F}$ inside the fluid volume $V^F$, i.e.
\begin{equation}
\rho= c \rho^A + \left( 1-c\right) \rho^W \qquad  {\rm with} \qquad
\mu= c \mu^A + \left( 1-c\right) \mu^W\; ,
\label{stoffgesetz}
\end{equation}
 where $ \rho^A, \rho^W$ and $ \mu^A, \mu^W$ refer to the bulk properties of the two fluids,  which are deemed to be constant in the present study. The employed equation governing the mixture fraction $c$ follows from the continuity equation (\ref{FluidVolConti}) and the immiscibility condition, i.e. $Dc/Dt=0$, 
\begin{equation}
\frac{\partial c}{\partial t}+\frac{\partial \left( c v^F_i \right)}{\partial x_i}=0\;.
\label{contifresco}
\end{equation}
 Mixture fraction values $c=0.5$ are used to identify the interface.

\section{Numerical method}
\label{NUM}
The present paper refers to a pressure-based Finite Volume (FV) formulation using a cell centered, co-located variable arrangement on unstructured polyhedral grids \cite{Ferziger}. Such frameworks are frequently employed in fluid engineering applications. Integrals are approximated using a second-order accurate mid-point integration rule. Implicit first-order time discretisations are used and convective fluxes are approximated by first-order upwind (UDS) baseline formulae and subjected to explicit deferred corrections for higher-order approximations. In the present study, a simple flux blending scheme featuring 70\% second-order central differencing are used. Diffusion fluxes are obtained from central differences, which also employ a deferred correction approach to account for non-orthogonality and face interpolation related issues.  
The procedure largely follows \cite{Ferziger}. Further details are given in \cite{Yakubov2015} and \cite{Svenja2017}. We firstly outline the governing system and its discretised implementation based upon $\hat{v}_i$ and $p^F$. Subsequently the approach is discussed for a formulation based upon $v^F_i$ and $p^F$.  

\subsection{\texorpdfstring{Governing equations using $\hat{v}_i$ and $p^F$.}{Governing equations for incompressible two-phase fluid flow through rigid porous media with variables \hat{v}\_i and p\textasciicircum F}} 
\label{GovEqVHatPF}
The employed continuity equation $\partial \hat v_i /\partial x_i=0$ (\ref{WholeVolConti}) has already been outlined in Sec. \ref{ConEq}. The momentum equation follows from equation \eqref{MomTheor}, which is in line with e.g. Del Jesus et. al \cite{DelJesus2012}, for a rigid inhomogenous porous material with exception of the diffusion term (see equation \eqref{XMomEq1a}). The difference in the diffusion term is attributed to our aim at obtaining a conservative form that serves a FV implementation. This conservative form follows from dividing equation \eqref{MomTheorCons} by $n$ which is valid for rigid porous media, i.e. 
\begin{equation}
 \frac{1}{n} \left[  \frac{\partial \; \rho \hat{v}_i}{\partial t}+  \frac{\partial}{\partial x_{j}} \left(\frac{\rho \hat{v}_j \hat{v}_i }{n}-\mu \frac{\partial \hat{v}_i}{\partial x_{j}}\right)
 \right] 
 - 
 \frac{\hat{v}_i}{n}
 \left[ 
 \frac{\partial \; \rho}{\partial t}+ 
 \frac{\partial}{\partial x_j}\left(\frac{\rho \hat{v}_j }{n} \right)
 \right]
 = -\frac{\partial  p^F}{\partial x_i}+\rho g_i-\frac{\rho}{n} \hat{v}_i \left( \tilde A  + \tilde B \mid \underline{ \hat{v}} \mid \right) \;.
 \label{MomEqVar1a}
\end{equation}
Note that to let the second bracket on the LHS vanish, the conservative air mixture fraction transport must employ the fluid velocity and not the Darcy velocity, cf. 
 (\ref{contifresco}), viz.
\begin{equation}
\frac{\partial c}{\partial t}+\frac{\partial \left( c \frac{\hat{v}_i}{n} \right)}{\partial x_i}=0\;.
\label{freesurfacedurchn}
\end{equation}

\subsubsection{\texorpdfstring{Discretised momentum equation}{Discretised momentum equation using variables \hat{v}\_i and p\textasciicircum F}} 
\label{MomEqDiscrVHatpF} 
\begin{figure}[htbp]
\centering
  \includegraphics{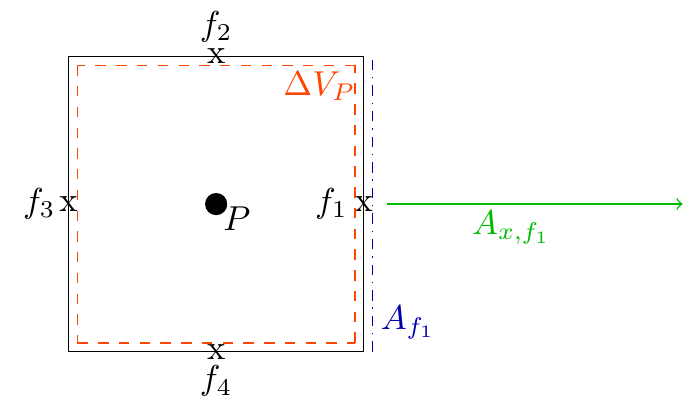}
  \caption{Schematic image of a finite volume cell.}
  \label{FVCellBild}
\end{figure}

We 
discretise Eqn. (\ref{MomEqVar1a}) over a control volume $\Delta V_P$ around its center $P$ as depicted in Figure \ref{FVCellBild}. Using a mid-point integration rule together with a simple first-order implicit time discretisation and an implicit upwind-difference scheme part for the convective flux yields
\begin{equation}
\begin{split}
\hat{v}^{n,m}_{i, P} \; \left[ \Delta V_P \, \left(\frac{\rho}{n}\right)_P \left( \frac{ 1}{\Delta t}+ \tilde A + \tilde B \mid \underline{\hat{v}}\mid^{n,m-1}   \right)_P 
+ \sum_{f(\Delta V_P)} A_{NB}
\right] \\.
- \sum_{f(\Delta V_P)}
\underbrace{\left[ \left( \frac{1}{n} \; max \left \lbrack -\frac{\dot{\hat m}^{n,m-1}}{n},0 \right \rbrack  \right)_f + \left(\frac{1}{n} \; \frac{\mu A} {d} \right)_f \right]}_{A_{NB}} \hat{v}^{n,m}_{i,NB}
=\\
-\sum_{f(\Delta V_P)}\left(p^{F,n,m-1}_{f}A_{i}\right)+
\rho_P \Delta V_P \left( g_i+ \frac{\hat{v}^{n-1}_{i}}{n \Delta t}
\right)_P +S_{\hat v_i} \;.
\label{MomDiscrHatVpF}
\end{split}
\end{equation}
Here $\dot {\hat m}_f = (\rho \hat v_i A_i)_f $ refers to porous mass flux across a face, the subscript $f$ marks  the cell faces and related variable values (see \ref{FVCellBild}), $A$ denotes the face area and $A_{i}$ the outward pointing face vector. The scalar distance between the cell center $P$ and adjacent neighbouring centers $NB$ is labeled $d$. For each time step, the governing equations are iterated to convergence in a segregated manner. The superscripts $n$ and $m$ denote to the time step and the iteration index of the iterative procedure, respectively. Porous resistance terms are implemented in an implicit manner which improves the stability. The source term $S_{\hat v_i}$ includes explicit terms which arise from different deferred correction contributions, e.g. higher-order convection, non-orthogonality and interpolation corrections.  

\subsubsection{\texorpdfstring{Discretised continuity equation}{Discretised pressure equation using the variables \hat{v}\_i and p\textasciicircum F}}
\label{SIMPLEErkl}
A SIMPLE-type pressure correction scheme is used to fulfill the continuity equation, cf. \cite{Ferziger}, \cite{Yakubov2015}. The procedure follows a fairly standardised approach and is only briefly outlined here. It starts from the values ($\hat{v}_{i}^{m-1}$, $p^{F,m-1}$) of the previous iteration state. An intermediate Darcy velocity $\hat{v}_{i}^*$ obtained from (\ref{MomDiscrHatVpF}) is usually not divergence free, therefore a velocity correction $\hat{v}^{'}_i: \hat{v}^{m}_i=\hat{v}^{*}_i+\hat{v}^{'}_i$, is introduced to satisfy the discrete continuity equation for $\hat{v}^{m}_i$. The  velocity correction is linked to a pressure correction $p': p^{F,m} = p^{F,m-1} + p^{'}$
using  a simplified momentum expression (\ref{MomDiscrHatVpF}), i.e. 
\begin{equation}
\hat{v}^{'}_{i}=-\frac{\Delta V}{A_{P}^{\hat{v}_i}}\frac{\partial p^{'}}{\partial x_i} \, , 
\label{vdash}
\end{equation}
where $A_{P}^{\hat{v}_{i}}$ is the main-diagonal coefficient of the  momentum equation (\ref{MomDiscrHatVpF}). Expression (\ref{vdash}) is substituted into the integral form of the continuity equation $\sum_f A_i (\hat{v}_{i}^{*} + \hat{v}_{i}^{'} ) = 0 $ and discretised to compute the pressure correction. Note that the second term of equation \eqref{contiVolV} is neglected at this point. Subsequently (\ref{vdash}) is used to determine $\hat{v}_{i}^{'}$. To avoid  pressure oscillations occurring for a co-located variable arrangement, a special interpolation of the fluxes in $(\hat{v}_i^* A_i)_f$ is used \cite{RhieChow}. It is important to note that the classical SIMPLE pressure correction scheme can be adopted without any change. Porous media influences are implicitly considered through the main diagonal coefficients $A_{P}^{\hat{v}^F_i}$, cf. (\ref{MomDiscrHatVpF}). 

\subsubsection{\texorpdfstring{Discretised air mixture fraction equation}{Discretised free surface equation using variables \hat{v}\_i and p\textasciicircum F}}
\label{EQFREES}
Using  a first-order  implicit time integration scheme and an implicit upwind-difference scheme part for the approximation of the convective term, the discrete mixture fraction transport reads
\begin{equation}
\begin{split} c^{n,m}_{P} \left[
\frac{\Delta V_P}{\Delta t} 
+\sum_{f(\Delta V_P)} A_{NB} \right] 
-\sum_{f(\Delta V_P)}
\underbrace{\left(max \left \lbrack - \frac{(\dot{\hat m}/\rho)}{n}^{n,m-1},0 \right \rbrack \right)_f}_{A_{NB}} c^{n,m}_{NB}
=\frac{\Delta V_P }{\Delta t}c^{n-1}_{P}+S_c \, , 
\label{DisplFreeSurfHatVpF}
\end{split}
\end{equation}
where $S_c$ hosts the  deferred correction terms of the compressive approximation. Regarding this, both the HRIC (e.g. \cite{Muzaferija1998}) scheme and the CICSAM (e.g \cite{Ubbink1997}, \cite{Ubbink1999}) scheme are used in the present study. With the air mixture fraction field, the values of the  cell-/face- centered density and viscosity are updated at each iteration by the cell-/face-centered values of $c$ using the simple linear equation of state (\ref{stoffgesetz}).

\subsection{\texorpdfstring{Governing equations using $v^F_i$ and $p^F$.}{Governing equations for incompressible two-phase fluid flow through rigid porous media with variables v\_i\textasciicircum F and p\textasciicircum F}} 
\label{GovEqVFPF}
In the second implementation, the fluid velocity $v^F_i$ and the fluid pressure $p^F$ are used as dependent variables. The continuity equation reads
\begin{equation}
\frac{\partial  v^F_i }{\partial x_i}=0 \, 
\label{ConFreeSurf}
\end{equation}
for a fluid volume $d V^F$. The momentum equation is again displayed in a conservative form (equation \eqref{MomTheorCons} divided by $n$ which is valid for rigid porous material)
\begin{equation}
  \frac{\partial \; \rho v^F_i}{\partial t}+ 
  \frac{\partial }{\partial x_j} \left(
  \rho v^F_j v^F_i - \mu \frac{\partial v^F_i}{\partial x_j}\right)
  - \left[ 
  \frac{\partial \; \rho}{\partial t} + v^F_i
  \frac{\partial \; \rho v^F_j}{\partial x_j} 
  \right] 
  =-\frac{\partial p^F}{\partial x_i}+\rho g_i-\rho v^F_i \left( \tilde A +  n \tilde B \mid \underline{v}^F \mid
  \right) \; ,
 \label{MomEqVar3}
\end{equation}
and a conservative air mixture fraction equation is used to close the system, viz. 
\begin{equation}
\frac{\partial c}{\partial t}+\frac{\partial \left( c v^F_i \right)}{\partial x_i}=0\;.
\end{equation}

\subsubsection{\texorpdfstring{Discretised momentum equation}{Discretised momentum equation using variables v\textasciicircum F\_i and p\textasciicircum F}} 
The governing equations should be integrated  over the fluid volume $\Delta V^F$, which is usually not explicitly available.  
 Using $dV^F = n dV$ and $dA_i^F= n dA_i$ etc., together with a midpoint integration rule, the aforementioned discretisation techniques yield 
\begin{equation}
\begin{split}
 {v}^{F,n,m}_{i, P} \;  \left[ \rho \Delta V_P \, \left( \frac{ 1}{\Delta t}+ \tilde A + \tilde B n \mid \underline{{v}}^F \mid^{n,m-1}   \right)_P 
+ \sum_{f(\Delta V_P)} A_{NB}
\right] \\.
- \sum_{f(\Delta V_P)}
\underbrace{\frac{n_f}{n_P}\left[ \left(max \left \lbrack -\dot m^{n,m-1}_f,0 \right \rbrack  \right)
+ \left(\frac{\mu^F A}{d}\right)
\right]}_{A_{NB}} v^{F,n,m}_{i,NB} = \\
-\sum_{f(\Delta V_P)}\frac{n_f}{n_P}\left(p^{F,n,m-1}_{f}A_{i}\right)+
\rho^F_P \Delta V_P \left( g_i+ \frac{{v}^{F,n-1}_{i}}{ \Delta t}
\right)_P +S_{v_i^F}\;,
\label{MomDiscrVFPFa}
\end{split}
\end{equation}
where $\dot m_f = (\rho v_i^F A_i)_f $ refers to a mass flux. The term $S_{v_i^F}$ formally coincides with the term in the previous momentum equation (\ref{MomDiscrHatVpF}). 
If the aim of the numerical method is to obtain conservative fluxes, which is usually the case for a FV approach, a unique porosity ratio $n_f/n_P$ (porosity value at face/ porosity value in cell center) is required in the flux terms. This suggests to interpolate the denominator $n_P$ to the faces, which immediately yields $n_{P_f} = n_f$ and the simpler expression
\begin{equation}
\begin{split}
 {v}^{F,n,m}_{i, P} \;  \left[ \rho \Delta V_P \, \left( \frac{ 1}{\Delta t}+ \tilde A + \tilde B n \mid \underline{{v}}^F \mid^{n,m-1}   \right)_P 
+ \sum_{f(\Delta V_P)} A_{NB}
\right] \\.
- \sum_{f(\Delta V_P)}
\underbrace{\left[ \left(max \left \lbrack -\dot m^{n,m-1}_f,0 \right \rbrack  \right)
+ \left(\frac{\mu^F A}{d}\right)
\right]}_{A_{NB}} v^{F,n,m}_{i,NB} = \\
-\sum_{f(\Delta V_P)}\left(p^{F,n,m-1}_{f}A_{i}\right)+
\rho^F_P \Delta V_P \left( g_i+ \frac{{v}^{F,n-1}_{i}}{\Delta t}
\right)_P +S_{v_i^F}\;,
\label{MomDiscrVFPFb}
\end{split}
\end{equation}

\subsubsection{\texorpdfstring{Discretised continuity equation}{Discretised pressure equation using variables v\_i\textasciicircum F and p\textasciicircum F}}
The same SIMPLE algorithm as described in \ref{SIMPLEErkl} is used to iterate the pressure. Since \eqref{ConFreeSurf} also has to be integrated over the fluid volume, the face areas occurring in the discretised version are augmented by the face-centered porosities $n_{f}$  and the semi discrete pressure correction equation reads 
\begin{equation}
\sum_{f(\Delta V_P)} \left( n v^{*F}_{i}A_{i}\right)_f -\sum_{f(\Delta V_P)} \left[ \left(\frac{n \, \Delta V}{A_{P}^{v^F_i}}\right) \left(\frac{\partial p^{'}}{\partial x_i}\right) A_{i}\right]_f =0\;.
\label{presscorrFluidN1}
\end{equation}
Note that (Rhie-Chow type) corrections employed to avoid pressure oscillations are also multiplied with $n_f$ in line with equation \eqref{presscorrFluidN1}.

\subsubsection{\texorpdfstring{Discretised air mixture fraction equation}{Discretised free surface equation using variables v\_i\textasciicircum F and p\textasciicircum F}}
 Discretisation techniques outlined above are also used for concentration equation, viz. 
\begin{equation}
\begin{split}
c^{n,m}_{P} \left[
\frac{\Delta V_P}{\Delta t} 
+\sum_{f(\Delta V_P)} A_{NB} \right] 
-\sum_{f(\Delta V_P)}
\underbrace{ \frac{n_f}{n_P}
\left( max \left \lbrack - (\dot m / \rho) ,0 \right \rbrack \right)_f }_{ A_{NB} } \; c^{n,m}_{NB}
=\frac{\Delta V_P }{\Delta t}c^{n-1}_{P}+S_{c}
\label{DisplFreeSurfvFpF}
\end{split}
\end{equation}
Mind that we assume $n_{P_f} = n_f$ in line with Eqn. (\ref{MomDiscrVFPFb}). Fluid properties are updated using $c$ and  the  equations of state \eqref{stoffgesetz}.

\section{Two-Dimensional Examples} 
\label{2DTestcases}
In this section different 2D test cases, i.e. variants of unsteady dam break flows and a steady embankment flow,   
are investigated using different porous materials. The cell-centered porosities ($n$) and the constants of a reduced Engelund law ($\tilde A$, $\tilde B$, $\tilde C \equiv 0$) are defined in line with data used by previous computational and  experimental work. Face values of the porosity follow from a linear interpolation. To estimate non-linear influences on the porous resistance, 
average and peak porous Reynolds number $Re_P$ are calculated from the present simulations. The unconstraint external flows feature a fairly low Reynolds number. Thus the fluid simulation model follows a laminar representation -- i.e. no turbulent eddy viscosity etc. is considered -- and potential turbulent effects inside the porous material are deemed to be incorporated in the porous resistance force model.  
Results of the present simulations utilize a compressive HRIC approximation of convective fluxes (\cite{Muzaferija1998}) in test cases \ref{Dambreak2D} and \ref{3Dtestcases} as well as the CICSAM scheme (\cite{Ubbink1997}) in test case \ref{AcaseExp}. An implicit first-order accurate scheme is used to approximate time derivatives. 

\subsection{Dam break flow through rigid porous material - Lin (1998)}
\label{Dambreak2D}
The first case refers to a dam break through porous material as proposed in \cite{Lin1998} and \cite{Liu1999}. Experimental data from \cite{Liu1999} is used to validate the results. 
\begin{figure}[htbp]
\centering
  \includegraphics{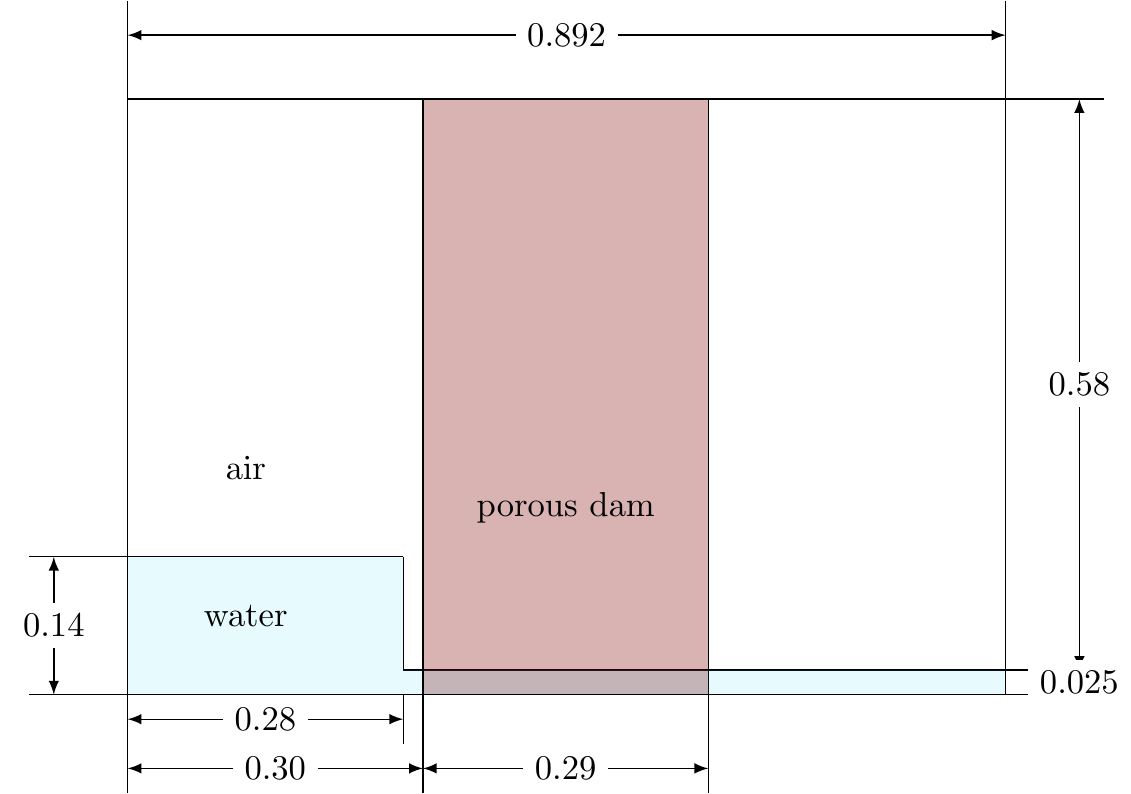}
  \caption{Geometry of porous dam break test case. All measurements are given in metres.} 
  \label{SetUpPorDamBreak}
\end{figure}The experiment features  a water column with an initial height of $h=14.0$\;cm and an initial width of $w=28.0$\;cm, which passes through a matrix of glass beads with an average diameter of  $D_{50}=0.3$\;cm that forms a porous material with a porosity value of $n=0.39$. The geometry is depicted by figure \ref{SetUpPorDamBreak}.
Using the present results, a low average Reynolds number of $Re_P=55$ follows from the space/time-mean Darcy velocity. The corresponding maximum  velocity yields a Reynolds number of $Re_P=125$. These values indicate, that nonlinear or turbulent effects should be of relevance for the resistance modelling. Following table \ref{PorForceVergl}, three different representations of the porous forces $\hat{f}^R_i$ are found in the literature. Larese et al.  \cite{Larese2015} use the Engelund equation, i.e. to determine the porous resistance from $\alpha=1000$ and $\beta=0.25$, which leads to  $\tilde  A=64.7 s^{-1}$ and $ \tilde B=334.2m^{-1}$. Similarly, Del Jesus et al. \cite{DelJesus2012} employ  $\alpha=700$ and $\beta=0.5$ and arrive at  $\tilde A=45.3s^{-1}$ and $\tilde B=668.4m^{-1}$. Moreover, Liu et al. \cite{Liu1999} employ the Van Gent model with $\alpha=200$ and $\beta=1.1$ according to table \ref{PorForceVergl}. Using a Keulegan-Carpenter number, an estimated characteristic time of $T=1s$,  as well as   $\alpha=200$ and $\beta=1.1$ yields 
$\tilde A= 67.3s^{-1}$, $\tilde B= 1470.5m^{-1}$ and an added mass parameter 
$\tilde C= 0.21$.
\\
Several aspects are investigated in the remainder of this subsection. An initial study investigates as reported in \ref{MeshStudyPorDamBreak}. Subsequently, influences related to different numerical methods and formulation differences discussed in section \ref{NUM} are scrutinized and outlined in  \ref{CompResNumMethH14}. The impact of the different resistance laws previously used for this test case is discussed in section  \ref{CompPorForces}. Section \ref{CrushedRocks} is devoted to a different material and  initial conditions, which lead to a higher porous Reynolds number. Finally, a simplified momentum equation is introduced in \ref{SimplEquation} and an error study using different artificial porous materials is conducted to discover the applicability range of the simplified equation.

\subsubsection{A resolution study}
\label{MeshStudyPorDamBreak}
\begin{figure}[]
\begin{center}
\subfigure[$t=0.0$\;$\si{\second}$]{ 
\includegraphics{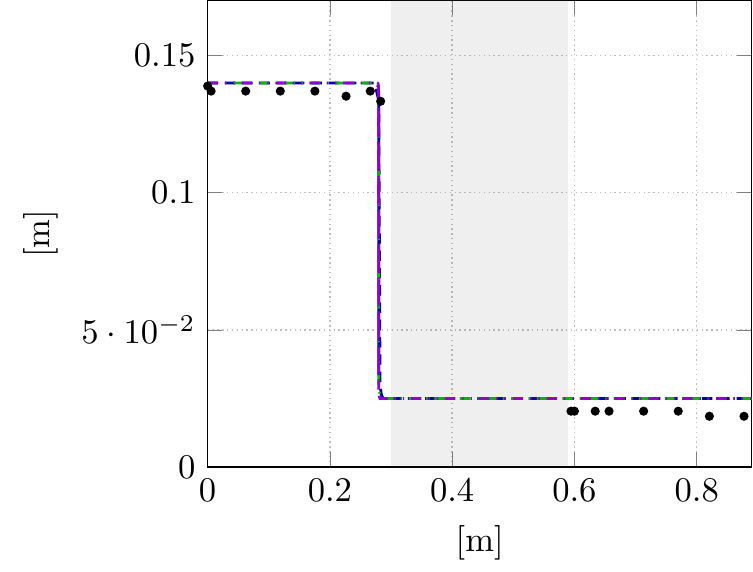}
}
\subfigure[$t=0.4$\;$\si{\second}$]{ 
\includegraphics{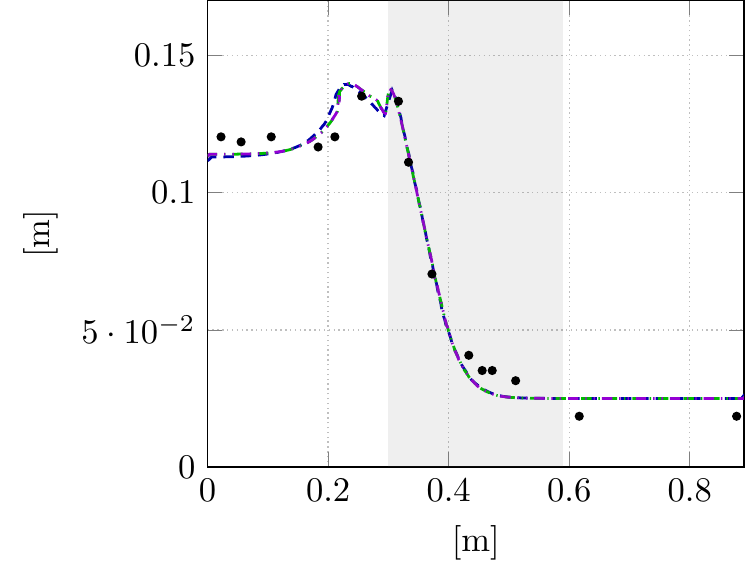}

}
\subfigure[$t=0.8$\;$\si{\second}$]{
\includegraphics{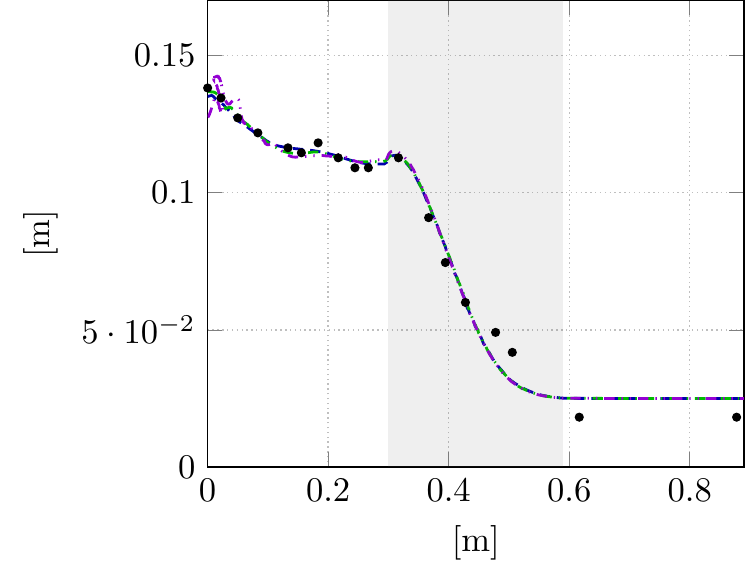}

}
\subfigure[$t=1.2$\;$\si{\second}$]{
\includegraphics{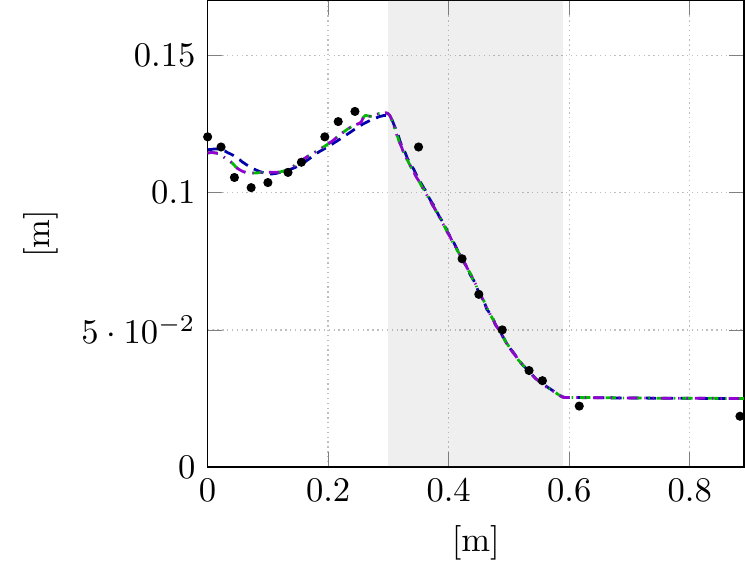}

}
\subfigure[$t=1.6$\;$\si{\second}$]{
\includegraphics{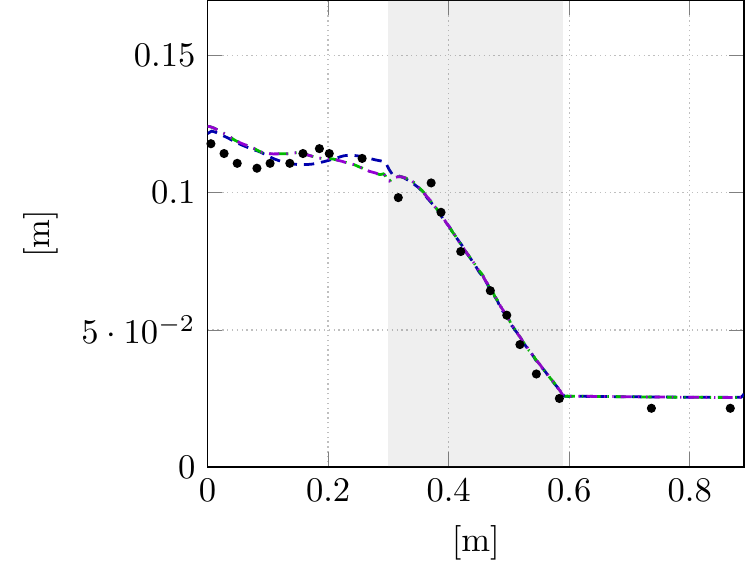}
}
\subfigure[$t=4.0$\;$\si{\second}$]{
\includegraphics{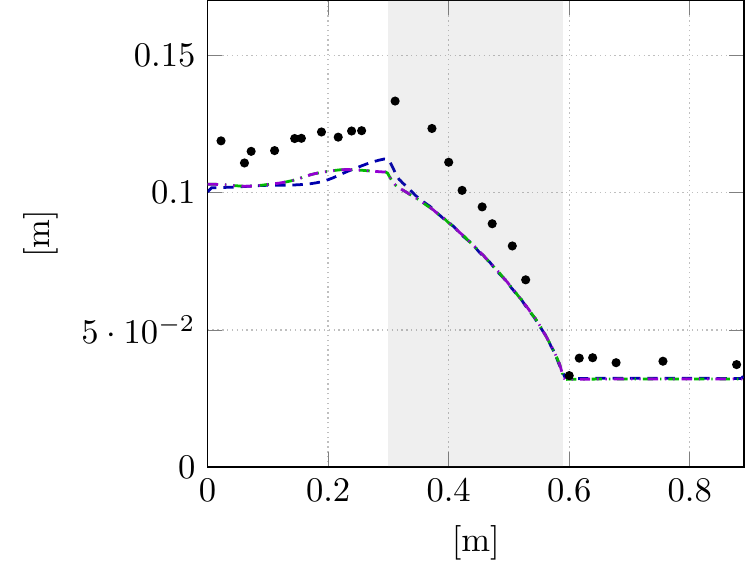}
}
\includegraphics{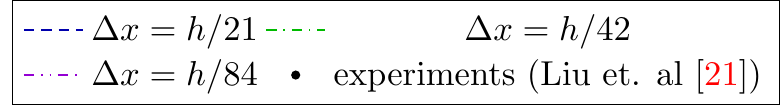}

\caption{Grid convergence study for the 
2D dam break test case ($h=14$\;$\si{\centi\metre}$) outlined by Fig. \ref{SetUpPorDamBreak} using the  $\hat{v}_i$-based VoF formulation. 
Comparison of the free surface evolution obtained from a 
coarse mesh ($\Delta x=h$/$21$),
medium mesh ($\Delta x=h$/$42$), 
 fine mesh ($\Delta x=h$/$84$) with 
experimental data of Liu et. al \cite{Liu1999}.} 
\label{MeshStudy_PorDamBreak_KorrVHatPF}
\end{center}
\end{figure}

 is conducted for all VoF implementations, using the Engelund equation with $\alpha=1000$ and $\beta=0.25$ ($\tilde A=64.7s^{-1}$, $\tilde B=334.2m^{-1}$). Simulations utilize three consecutively refined  isotropic homogeneous grids that feature $h/\Delta x= 21$, $h/\Delta x= 42$ and $h/\Delta x= 84$, which approximately refers to $\Delta x=0.665$\;$\rm{cm}$, $\Delta x=0.332$\;$\rm{cm}$ and $\Delta x=0.166$\;$\rm{cm}$. The employed time step sizes read $\Delta t= 2\cdot 10^{-4}$\;s, $\Delta t= 1\cdot 10^{-4}$\;s and $\Delta t= 5\cdot 10^{-5}$\;s  and 
ensured a Courant number well below $10^{-2}$. Exemplary results for the implementation method using the $\hat{v}_i$-based formulation can be seen in Figure \ref{MeshStudy_PorDamBreak_KorrVHatPF}. The results display mesh convergence for all displayed time instants. At $t=0.8$\;$\si{\second}$, the finest mesh resolves breaking wave phenomena for a reflected wave which is not seen by the coarser meshes. Since the medium grid is sufficient to have mesh convergence, it was selected for the following simulations.

\subsubsection{Modelling influences}
\label{CompResNumMethH14}
\begin{figure}[]
\begin{center}
\subfigure[$t=0.0$\;$\si{\second}$]{ 
\includegraphics{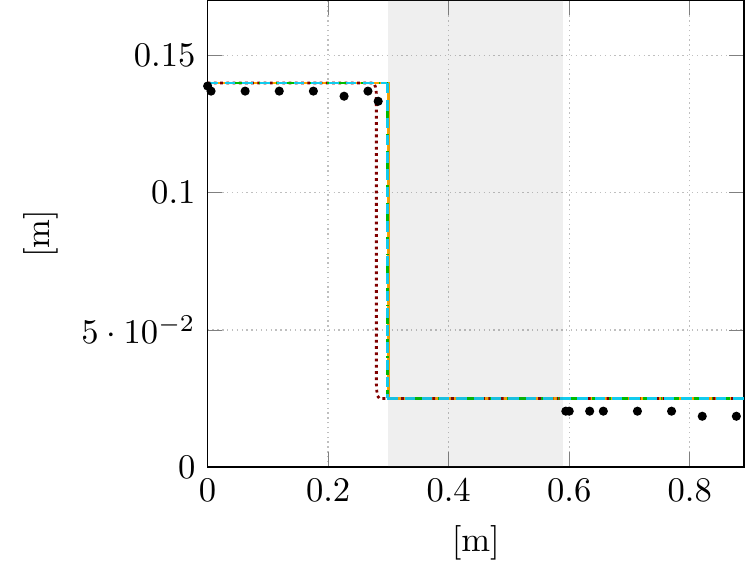}
}
\subfigure[$t=0.4$\;$\si{\second}$]{ 
\includegraphics{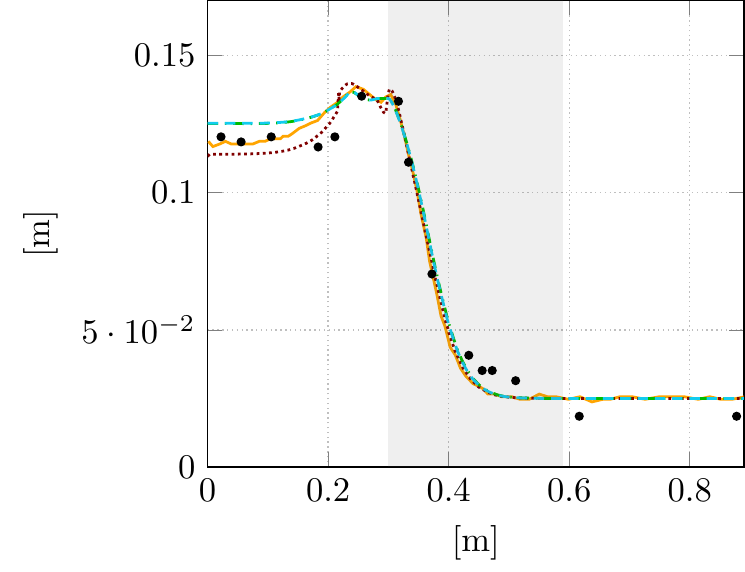}
}
\subfigure[$t=0.8$\;$\si{\second}$]{
\includegraphics{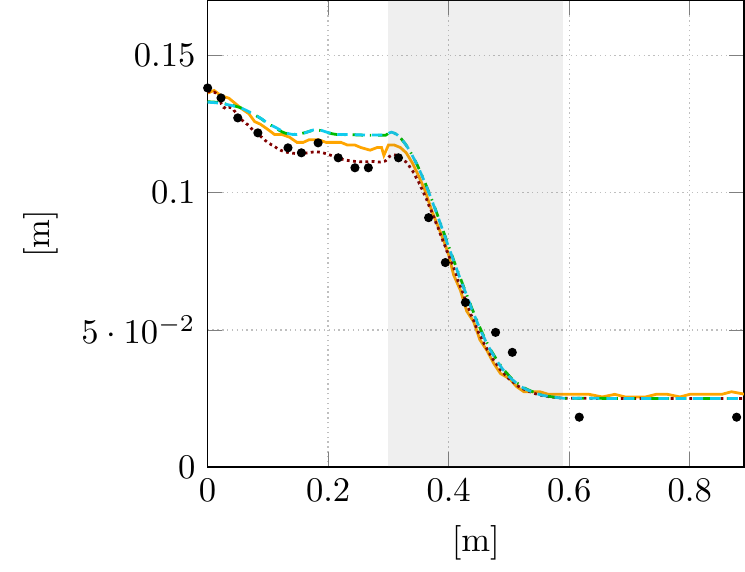}
}
\subfigure[$t=1.2$\;$\si{\second}$]{
\includegraphics{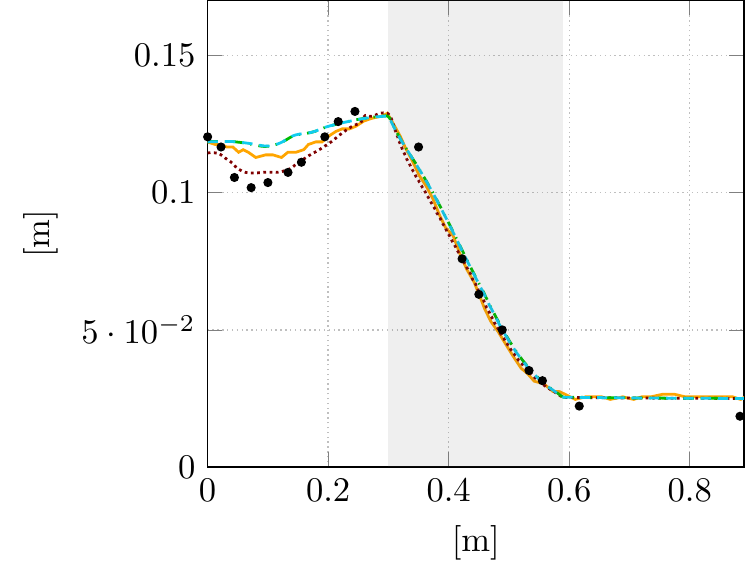}
}
\subfigure[$t=1.6$\;$\si{\second}$]{
\includegraphics{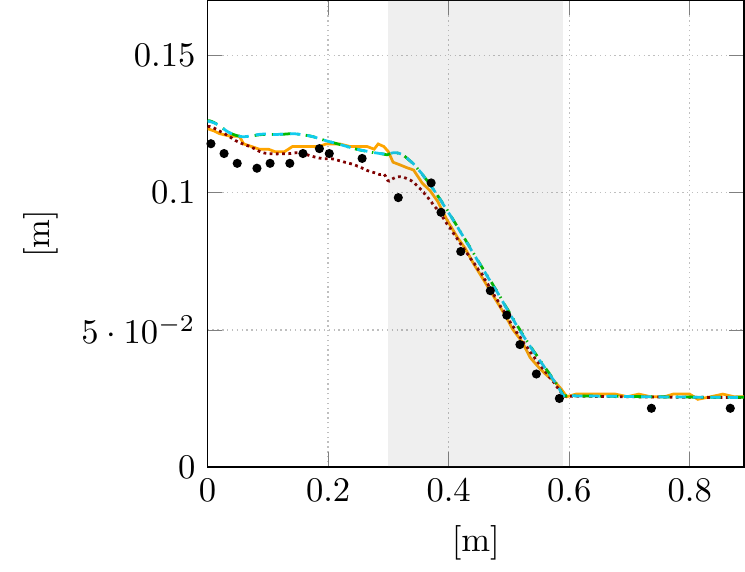}
}
\subfigure[$t=4.0$\;$\si{\second}$]{
\includegraphics{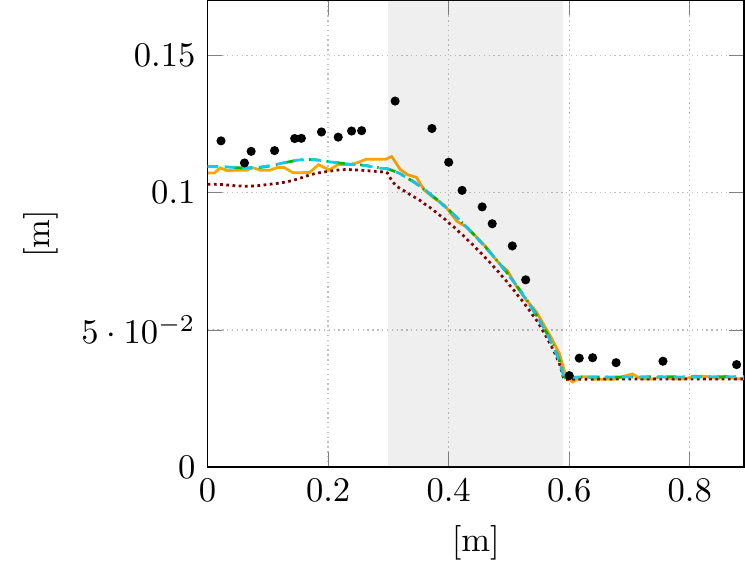}
}
\includegraphics{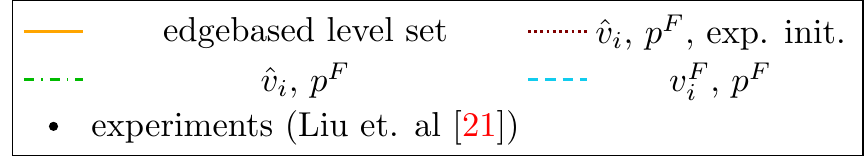}
\caption{Numerical modelling influences on the evolution of free surface for the 2D dam break test case ($h=14$\;$\si{\centi\metre}$). Results obtained from the edge-based level set solver 
, a $\hat{v}_i$-based VoF formulation for $w=29.8$ cm 
and $w=28$ cm 
,  and a $v^F_i$-based VoF formulation 
 compared with experiments of Liu et. al \cite{Liu1999}
. }
\label{PorDamBreak_DiffImplVersVsLarese}
\end{center}
\end{figure}

are outlined by Figure \ref{PorDamBreak_DiffImplVersVsLarese}, which compares experimental data  for the free surface evolution published in \cite{Liu1999} with numerical results obtained from different computational models. Simulations refer to an edge-based level-set solver reported by Larese et al \cite{Larese2015} and the two VoF formulations on the medium mesh as discussed in Sect. \ref{NUM}. All 
simulations employ the resistance law already used in the previous section \ref{MeshStudyPorDamBreak}.\\
Note that Larese et al. \cite{Larese2015} increased the experimentally reported initial column width from $w=28$ cm to $w=29.8$ cm. The increased width was also used by the present VoF simulations, to support the verification and assess formulation related differences. Moreover, we supplement FV results obtained from the $\hat{v}_i$-based formulation for the experimental  column width $w=28$ cm. No visible differences can be detected for the results obtained from the two VoF approaches, i.e. using the $\hat{v}_i$-based or $v^F_i$-based formulation. Comparing the results for the two different initial column widths predicted by the same FV formulation, it is seen that the correct initialisation provides a significantly better agreement with experimental data for all time instants except $t=4.0$\;s. A possible reason for this is described by Liu et al.  \cite{Liu1999}, as the free surface tends to adhere to the lateral glass wall. Therefore, the experimental data might  overestimate the free surface elevation at the late time instant.
Figure (\ref{PorDamBreak_DiffImplVersVsLarese}) also supports a comparison of different numerical methods using the same initial conditions and similar resolutions. Mind that Larese et al.  \cite{Larese2015} did use a level set approach which is restricted to the water phase whereas the present approach resolves both fluid phases using VoF.
Albeit the different governing equations, the results of the different solvers are in acceptable agreement for all time instants. Noticeable differences occur upstream the porous dam, and at the entrance of the water into the porous material. The transit of free surface elevation predicted by the FV solver is more continuous at $t=0.8$\;s, $t=1.6$\;s and $t=4.0$\;s, whereas slightly more damming is observed in the results of \cite{Larese2015}. Since the convection term of both formulations differs along the porous dam,  we attribute these predictive differences to the different formulations. However, slight damming due to dynamic effects also appears for the FV study when changing to the experimentally reported initialisation, which reveals a small gap of air between the liquid dam and the porous medium. It is noteworthy that the differences between the FV VoF solver and the edge-based level set solver are far more considerable than the differences between the two VoF formulations described in section \ref{NUM}. 

\subsubsection{Porous force model influences}
\label{CompPorForces}
\begin{figure}[]
\begin{center}
\subfigure[$t=0.0$\;$\si{\second}$]{ 
\includegraphics{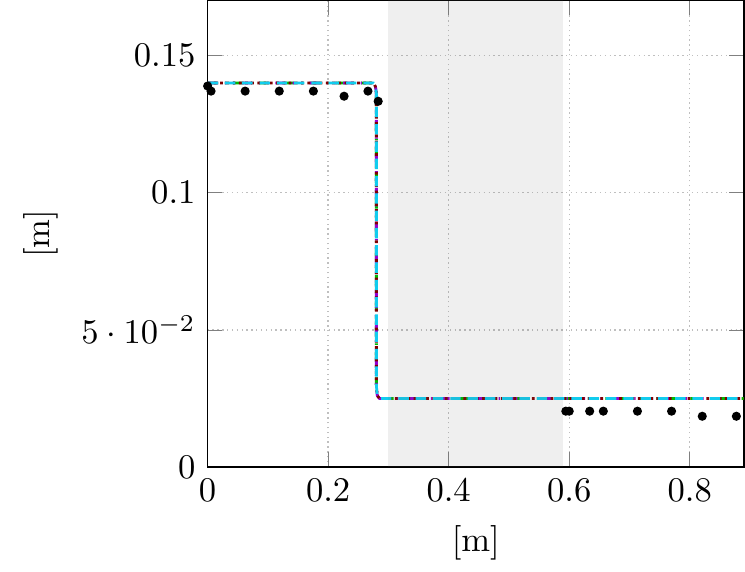}
}
\subfigure[$t=0.4$\;$\si{\second}$]{ 
\includegraphics{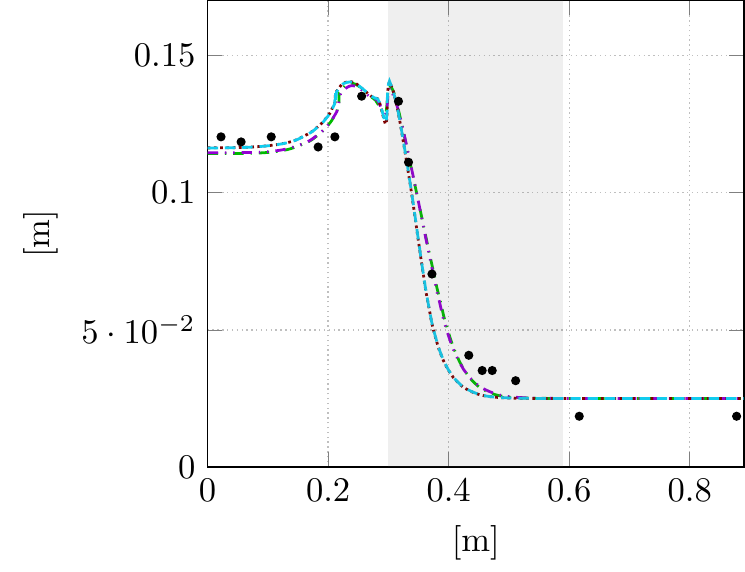}
}
\subfigure[$t=0.8$\;$\si{\second}$]{
\includegraphics{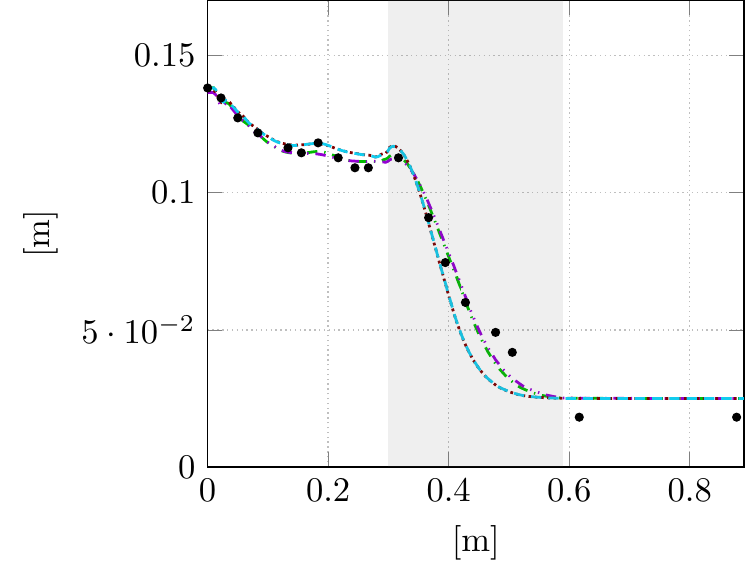}
}
\subfigure[$t=1.2$\;$\si{\second}$]{
\includegraphics{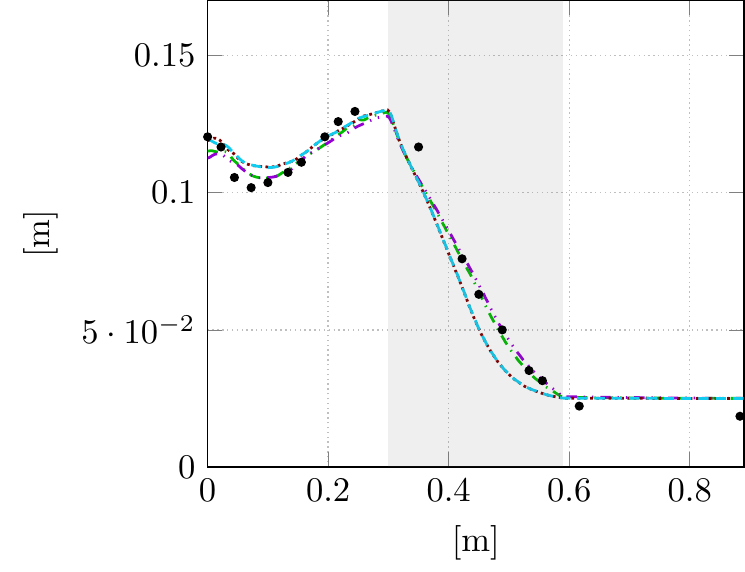}
}
\subfigure[$t=1.6$\;$\si{\second}$]{
\includegraphics{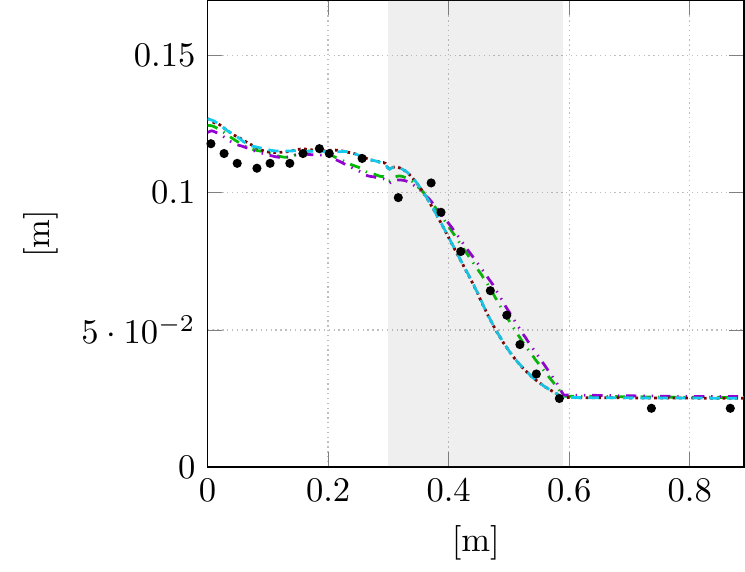}
}
\subfigure[$t=4.0$\;$\si{\second}$]{
\includegraphics{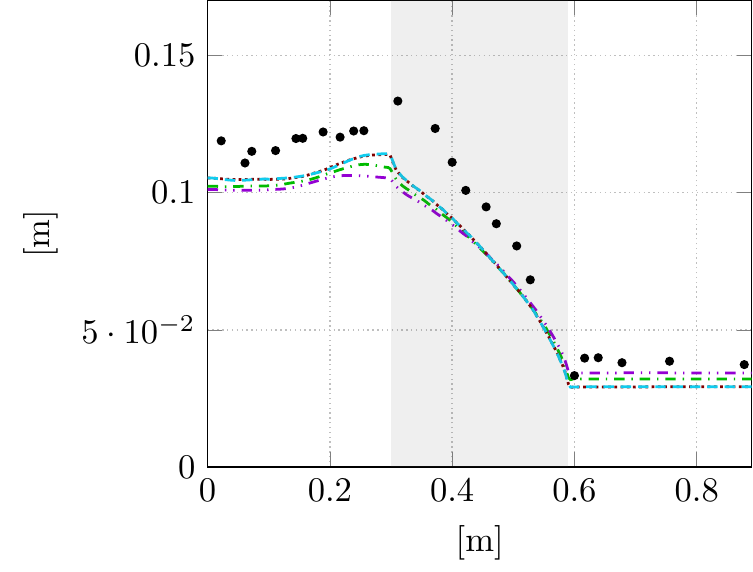}
}
\includegraphics{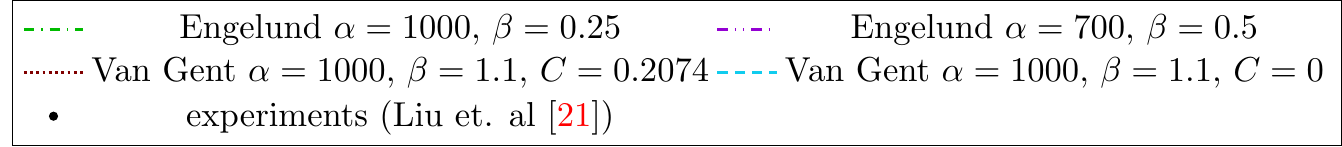}
\caption{Porous force modelling influences on the evolution of free surface for the 2D dam break test case ($h=14$\;$\si{\centi\metre}$). Results obtained from the $\hat{v}_i$-based VoF formulation for two Engelund models ($\alpha=1000$, $\beta=0.25$ 
; $\alpha=700$, $\beta=0.5$  
) and two Van Gent models ($\alpha=200$, $\beta=1.1$, $C=0.2074$  
; $\alpha=200$, $\beta=1.1$, $C=0$) 
 in comparison with experiments of Liu et. al \cite{Liu1999} 
 .}
\label{PorDamBreak_DiffResForm}
\end{center}
\end{figure}

are displayed in Figure \ref{PorDamBreak_DiffResForm} for FV simulations using the $\hat{v}_i$-based VoF formulation in conjunction with the experimentally reported initial conditions and the medium mesh described in Sec. \ref{MeshStudyPorDamBreak}. As indicated by the two Van Gent results, the 
influence of the added mass coefficient is negligible, which  explains why it was hardly considered by other authors. In contrast to the formulation and modelling  differences assessed  above, changes to the resistance law have a large effect on the development of the free surface. This highlights the dominance of the resistance law to balance pressure and gravity forces inside the momentum equation. The resistance law fitting best to the experimental data is the one used by Larese et al.  \cite{Larese2015} which was also used in the mesh study (cf. Sec. \ref{MeshStudyPorDamBreak}). 

\subsubsection{\texorpdfstring{Porous Reynolds number influences}{Crushed Rocks (h=25\;cm)}}
\label{CrushedRocks}
were studied by a 2D dam break case which involves a higher porosity index and a different initial water column. Modifications were extracted from the crushed rocks case described in \cite{Liu1999}. 
Results aim to assess predictive differences revealed by the two formulations discussed in section \ref{NUM} for a higher porous Reynolds number. The initial water column features a height of $25$\;cm and a width of $30$\;cm. The porous material consists of crushed rocks with a porosity index of $n=0.49$ and an average diameter $0.0159$\;m.
The case leads to $Re_P=1267$ for the space/time-averaged Darcy velocity and  $Re_P=3473$ for the maximum Darcy velocity observed in space and time. Therefore, a turbulent Van Gent \cite{VanGent1992} resistance model based upon $\alpha = 1000$ and $\beta = 1.1$ ($\tilde A=12.87$\;1/$\si{\second}$, $\tilde B=146.95$\;1/$\si{\metre}$) was employed. The added mass coefficient $\tilde C$ was neglected. Results are again assessed by the temporal evolution of the free surface and compared with experimental data reported by Liu et al. \cite{Liu1999}. \\
\begin{figure}[]
\begin{center}
\subfigure[$t=0.0$\;$\si{\second}$]{ 
\includegraphics{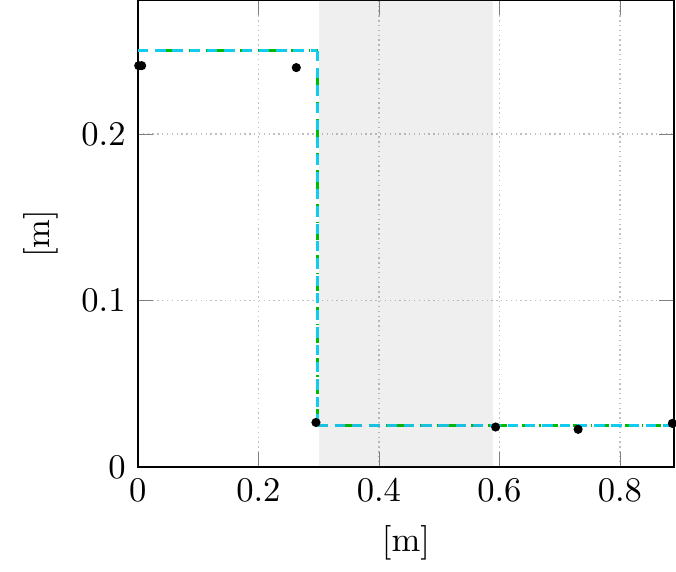}
}
\subfigure[$t=0.2$\;$\si{\second}$]{ 
\includegraphics{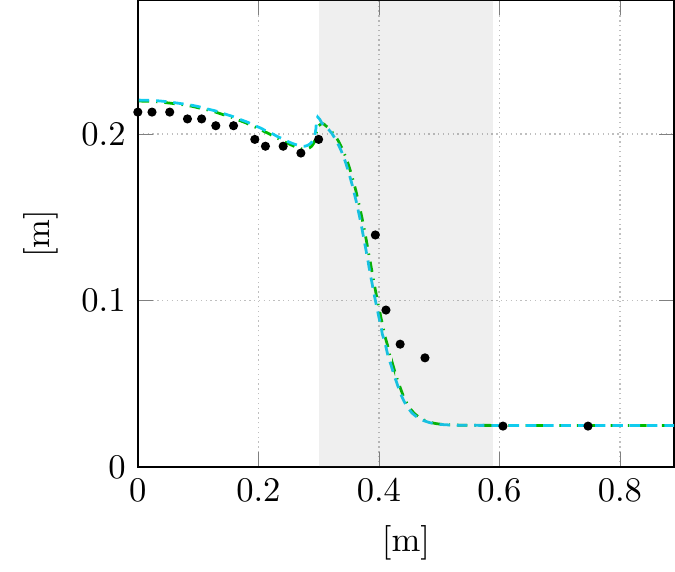}
}
\subfigure[$t=0.6$\;$\si{\second}$]{
\includegraphics{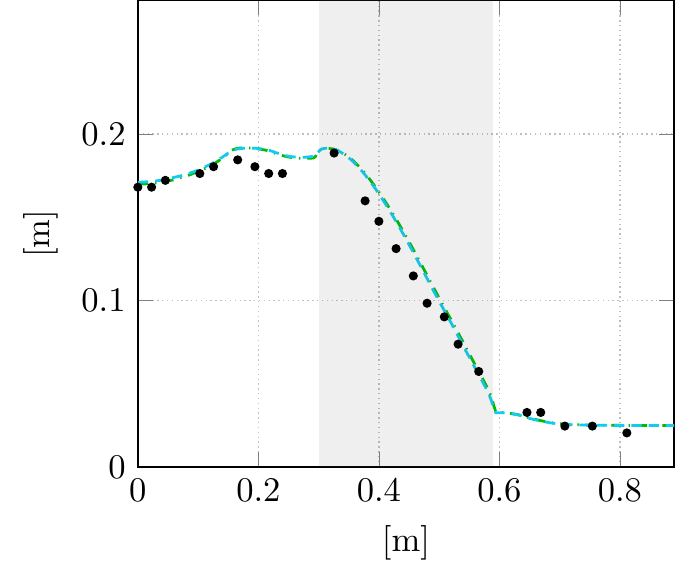}
}
\subfigure[$t=1.0$\;$\si{\second}$]{
\includegraphics{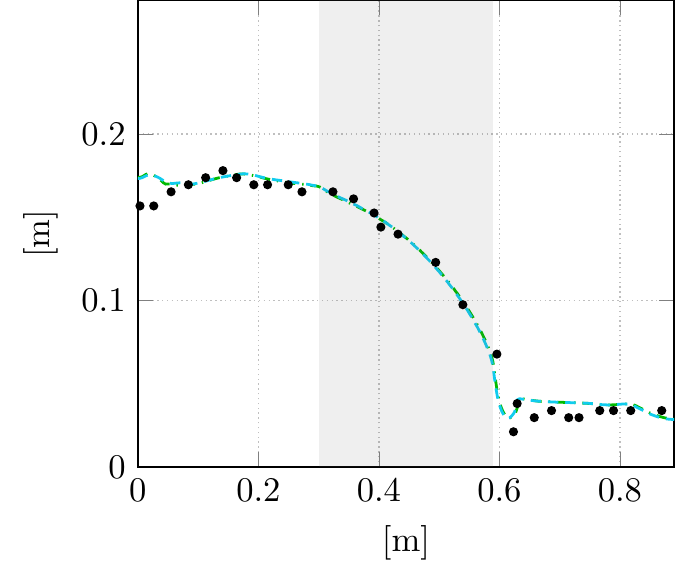}
}
\subfigure[$t=1.4$\;$\si{\second}$]{ 
\includegraphics{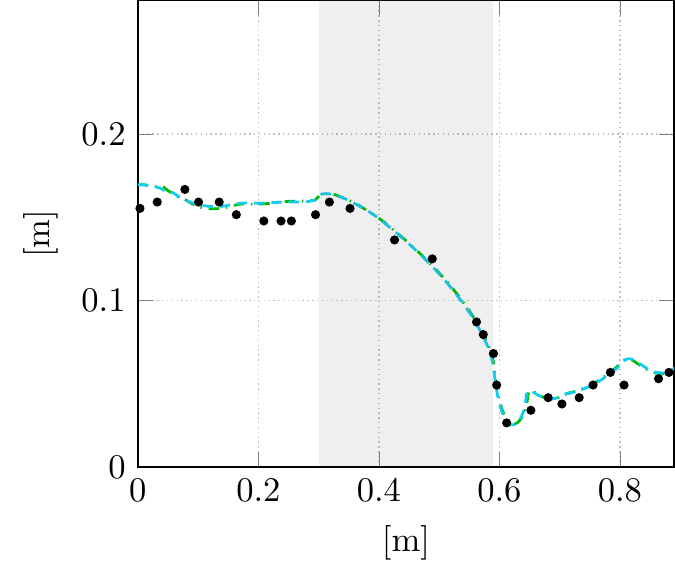}
}
\subfigure[$t=2.2$\;$\si{\second}$]{
\includegraphics{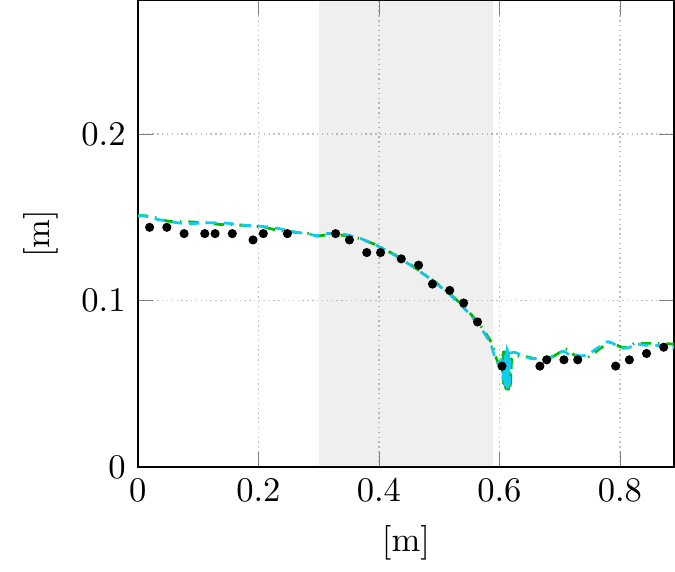}
}
\includegraphics{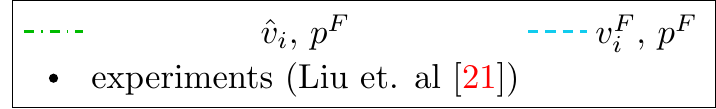}
\caption{Numerical modelling influences on the evolution of free surface for the 2D crushed rocks case ($h=25$\;$\si{\centi\metre}$). Evolution of the free surface predicted by the $\hat{v}_i$-based VoF formulation 
and the $v^F_i$-based VoF formulation 
in comparison with experiments of Liu et. al \cite{Liu1999} 
.}
\label{PorDamBreak_CrushedStones}
\end{center}
\end{figure}Figure \ref{PorDamBreak_CrushedStones} reveals that the results returned by the two VoF formulations discussed in section \ref{NUM} are only marginally different and a fair predictive agreement can be observed. Particularly, the elevation dip observed downstream the porous layer  from $t=1.0 -1.6$\;$\si{\second}$ is represented well. Small deviations between the two simulations and between simulated and measured data can 
 be observed inside the porous material for $t\le0.8$\;s. The numerical modelling influences are thus slightly more pronounced for the higher porous Reynolds number but still remain small. 

 \subsubsection{Introducing a simplified momentum equation} 
\begin{figure}[htbp]
\begin{center}
\includegraphics{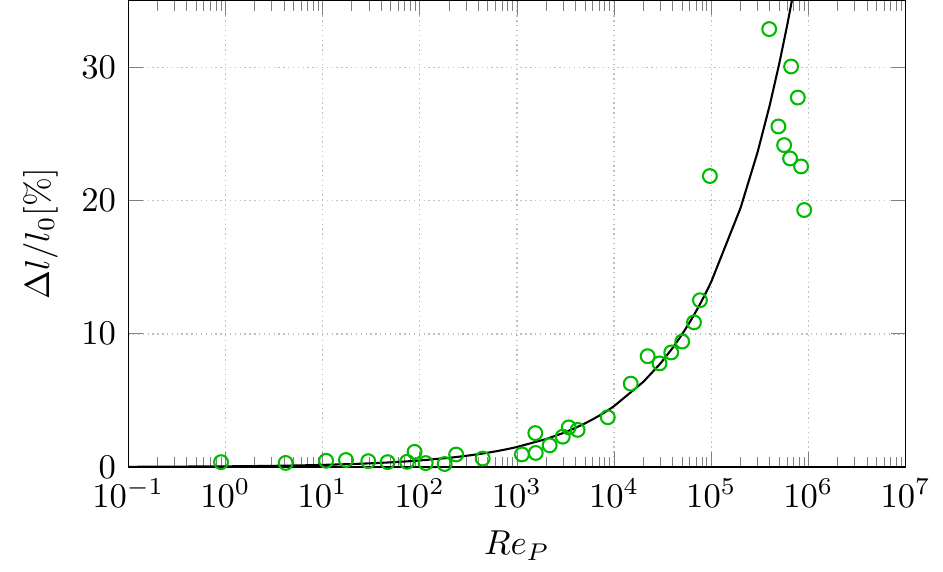}
\caption{Maximum error $\Delta l$/$l_0$ plotted over maximum Reynolds number.}
\label{ParamStudyPorDam_ErrVsReMeanMean}
\end{center}
\end{figure}

and studying its deviation from the more accurate framework \ref{GovEqVHatPF} 
 \label{SimplEquation}
 as a function of the porous Reynolds number provides an estimation of the errors due to the neglect of variable porosity influences inside the transport terms. 
 Results obtained in the  previous subsections suggest that the influence of transport terms is limited, whilst the pressure, volume and drag forces on the right hand side of the momentum equation     \eqref{MomEqVar1a} balance each other towards an equilibrium. 
 Therefore, the errors due to modifications of the transport terms might be limited. 
 The goal is thus to obtain an approximate correlation between the related error and the porous Reynolds number, which can be used to justify a simpler approach for a particular porous Reynolds number regime. 
 To this extent, simulations were conducted with the Darcy velocity using a simplified momentum equation  
 \begin{equation}
 \left[  \frac{\partial \; \rho \hat{v}_i}{\partial t}+  \frac{\partial}{\partial x_{j}} \left(\rho \hat{v}_j \hat{v}_i -\mu \frac{\partial \hat{v}_i}{\partial x_{j}}\right)
 \right] 
 - \hat{v}_i
 \left[  \frac{\partial \; \rho}{\partial t}+ 
 \frac{\partial}{\partial x_j}\left(\rho \hat{v}_j \right)
 \right]
 = -\frac{\partial  p^F}{\partial x_i}+\rho g_i-\frac{\rho}{n} \hat{v}_i \left( \tilde A  + \tilde B \mid \underline{ \hat{v}} \mid \right) \;.
 \label{eq:simpleM}
 \end{equation}
 For continuity reasons, the term inside the second square bracket 
 of \eqref{eq:simpleM} vanishes and the remaining expression features the usual conservative fluid dynamic transport term on the LHS together with the RHS force terms 
 that occur in the  Darcy velocity based momentum equation \eqref{MomEqVar1a}. Equation \eqref{eq:simpleM} denotes a traditional conservative fluid dynamic formulation which is only augmented by a porous force term on the RHS. The approach is therefore of interest, when either parts of the code are not accessible or implementation efforts should be kept small. Test case \ref{Dambreak2D} has been investigated with $36$ different porous media properties in the range of  $1 \le Re_p \le 10^{6}$. Resistance forces were derived from the Engelund equation found in table \ref{PorForceVergl} depending on the properties of the porous material together with $\alpha=1000$ and $\beta=1$. To obtain an integral error indication $E_{c^A}$ for the deviation of the simplified method from the full momentum equation described in \ref{GovEqVHatPF}, the difference between the air concentration fields of the simplified $c^A_{simpl}$ and the correct method $c^A_{corr}$ are spatially averaged 
\begin{equation}
E_{c^A}=\frac{\sum_{\Delta V_P}\mid c^A_{corr}-c^A_{simpl}\mid \Delta V_P}{\sum_{\Delta V_P} \Delta V_P}\;.
\label{EcA}
\end{equation}
As regards the transient behaviour of the spatially averaged indicator (\ref{EcA}), only the maxima of $E_{c^A}$ during a simulation time of $4.0$\;$s$ were employed. To obtain an error measure which represents the deviation of the dam shape and not the error of the concentration field (classically equation \eqref{EcA} non-dimensionalised with $c_{corr}^A$), a relation between the field error $E_{c^A}$ and a length based error $\Delta l$/$l_0$ is estimated. Here $\Delta l$ is the difference between the water column length $l$ and the initial water column length $l_0=0.28$\;m, i.e. $l_0+\Delta l=l$. The height $h$ of the shifted water column is calculated with the help of mass conservation. To obtain the relation of $E_{c^A}$ to $\Delta l$/$l_0$, $E_{c^A}$ is calculated for six artificially shifted geometries ($c^A_{simpl}$) in regards to the initial condition ($c_{corr}^A$). For all six geometry comparisons, the relation is found to be $\Delta l$/$l_0 \approx 1000 E_{c^A}$. Plotting the error measure $\Delta l$/$l_0$ over the maximum porous Reynolds number  in Figure \ref{ParamStudyPorDam_ErrVsReMeanMean}. 
a clear dependency of the deviation on the porous Reynolds number is observed. This indicates, that for 
small enough pore Reynolds numbers, the left-hand side terms are obviously much smaller than the right hand side terms which in turn seem to equilibrate. 
Using a nonlinear least-squares (NLLS) Marquardt-Levenberg algorithm for acquiring a fit leads to
\begin{equation}
E^{est., Re_P}_{c^A}=(0.00233211 Re_P)^{0.482901}\;.
\label{esterrorRe}
\end{equation}
For porous Reynolds numbers lower than $10^4$, the maximum deviation of results obtained from the simplified momentum equation (\ref{eq:simpleM}) falls below $5\%$. This motivates a possibility to simplify the momentum equation for all porous materials with mean grain diameters below $1$\;cm.

\begin{figure}[]
\begin{center}
\subfigure[$t=0.8$\;$\si{\second}$]{
\includegraphics{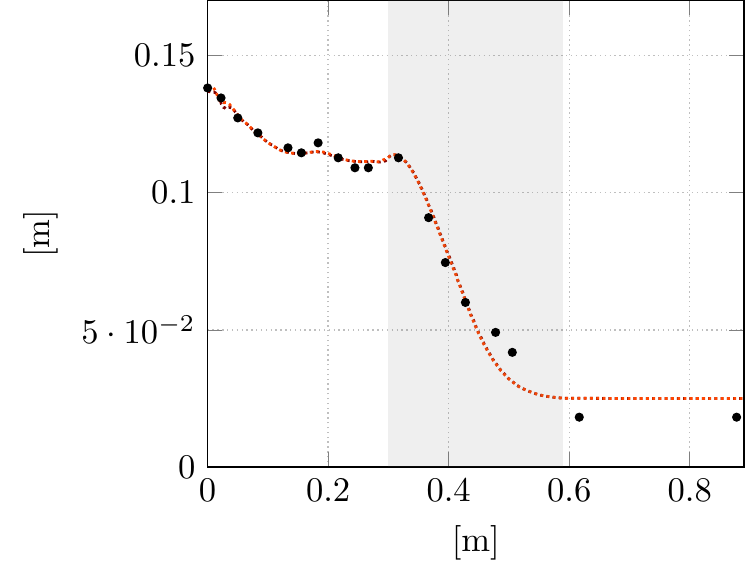}
}
\subfigure[$t=1.2$\;$\si{\second}$]{
\includegraphics{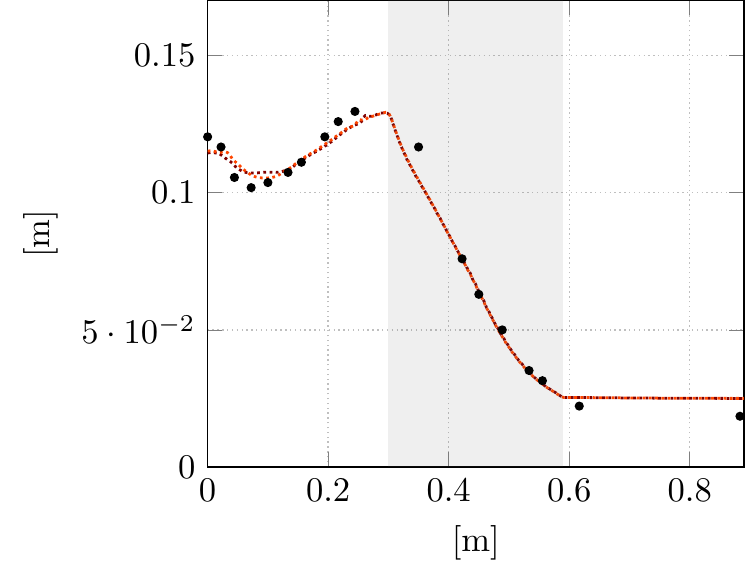}
}
\includegraphics{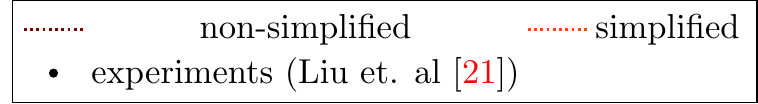}
\caption{Numerical modelling influences on the evolution of free surface for the 2D dam break test case ($h=14$\;$\si{\centi\metre}$). Results obtained from the simplified momentum equation \eqref{eq:simpleM} 
compared with results from non-simplified momentum equation 
with experiments of Liu et. al \cite{Liu1999} 
. }
\label{PorDamBreak_SimplVersVsExp}
\end{center}
\end{figure}

\subsubsection{The neglect of porous resistance forces} 
\label{extreme case}

\begin{figure}[]
\centering
\subfigure[$\hat{v}_x$]{
\includegraphics[scale=.35]{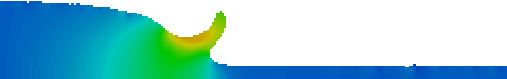}
}
\subfigure[$v^F_x$]{
\includegraphics[scale=.35]{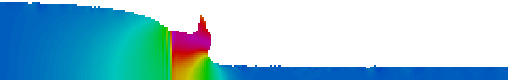}
}
\subfigure[]{
\includegraphics[scale=.35]{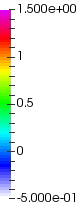}
}
\caption{Snapshot of the velocity coloured liquid body  predicted by the $\hat v_i$-based (left, Sec. \ref{GovEqVHatPF}) and the ${v}^F_i$-based (right, Sec. \ref{GovEqVFPF}) VoF formulations.}
\label{VelocityFieldsDarcy}
\end{figure}
  provides an illustrative example to  emphasize the expected differences of the two formulations described in Sec. \ref{NUM}.  
 A snapshot of the liquid body obtained from simulations without porous resistance,  
 i.e. $\tilde A=0, \tilde B=0$ and $\tilde C=0$, is displayed in Fig. \ref{VelocityFieldsDarcy}.
 The simulated liquid body is coloured by the Darcy velocity for the $\hat{v}_i,$-based formulation and fluid velocity for the $v^F_i$-based  formulation. As expected by definition, the fluid velocity strongly accelerates in the porous medium without porous resistance due to the reduction of the wetted regime and decelerates when leaving the porous media. On the contrary, the Darcy velocity displays a more continuous evolution over the fluid domain. Therefore, the $\hat{v}_i$-based formulation yields a smoother velocity field in the transition region, whereas the ${v}^F_i$-based approach returns large gradients that even yield slight oscillatory velocities in the transition region. This might relate to the assumption $A^F=n A$ in the continuity equation. In the absence of the usually dominating porous resistances, the formulation differences related to the transport terms lead to a visible difference of the free surface elevation. 
\begin{figure}[]
\centering
\subfigure[$v^F_x$]{
\includegraphics[scale=.38]{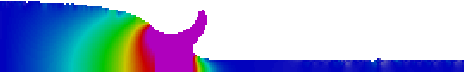}
}
\subfigure[$v^F_x$]{
\includegraphics[scale=.38]{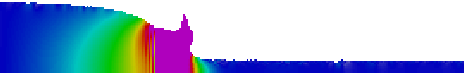}
}
\subfigure[]{
\includegraphics[scale=.35]{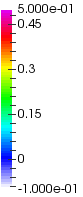}
}
\caption{Snapshot of the velocity coloured liquid body  predicted by the 
$\hat v_i$-based (left, Sec. \ref{GovEqVHatPF}) and the ${v}^F_i$-based (right, Sec. \ref{GovEqVFPF}) VoF formulations. }
\label{VelocityFieldsDarcy}
\end{figure}

\subsection{Flow through an inclined porous dam}
\label{AcaseExp}
\begin{figure}[htbp]
\centering
  \includegraphics{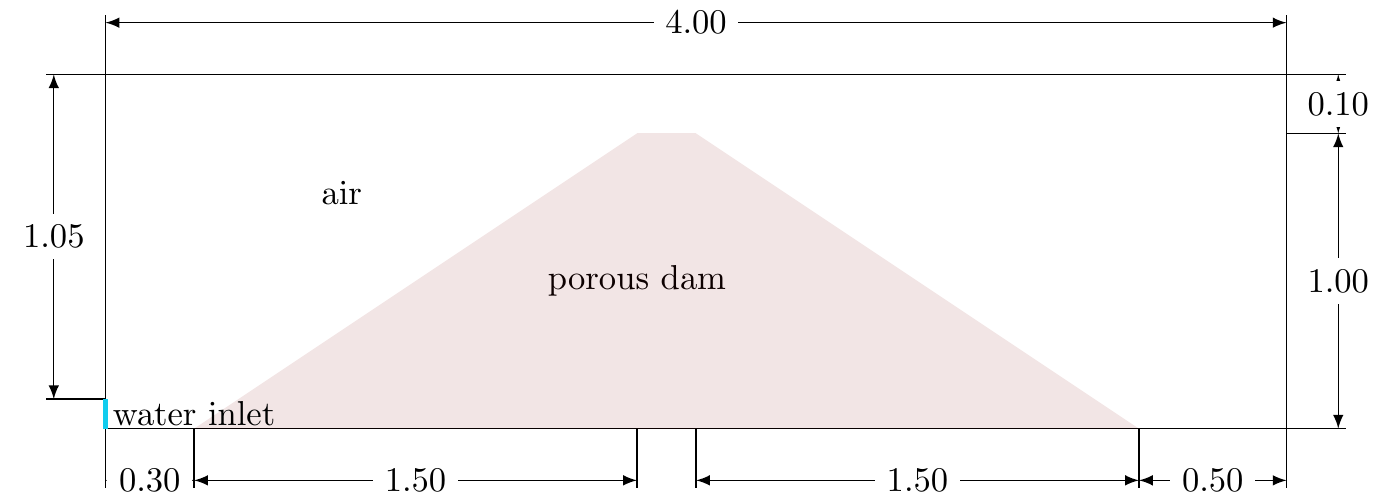}
 \caption{Geometry of the inclined homogeneous porous dam case.}
 \label{SetUpPorAcase}
\end{figure}

The second case refers to 2D experiments of water flow through an inclined dam which consists of homogeneous rockfill material as described in \cite{Larese2011}.
The selected case is illustrated in Fig. \ref{SetUpPorAcase}. It refers a constant volume flux of $Q=25.46$\;l/s entering a channel through an orifice of height $5$\;cm $\times$ width $246$\;cm, in which a dam of rockfill material is present. The porosity index of the dam reads $n=0.4052$ and the average particle diameter refers to $D_{50}=35.04$\;mm.
 This leads to a (computed) Reynolds number of $Re_P=1335$ using a space/time-averaged Darcy velocity and $Re_P=4939$ using the maximum Darcy velocity. The Ergun theory is used for the resistance law by means of 
\begin{equation}
\tilde A=150 \frac{\left(1-n\right)^2}{n^2}\frac{\mu}{\rho D_{50}^2}=0.26\;1/s
\end{equation}
\begin{equation}
\tilde B=1.75 \frac{\left(1-n\right)}{n^2} \frac{1}{D_{50}}=180.93\;1/m\;.
\end{equation}
\\ 
In accordance with the experiments, a Dirichlet condition using a homogeneous velocity $v^F_1=0.207$\;m/s is used to simulate the inflow and zero-gradient conditions were used at the outlet. The size of the considered 2D domain is indicated by Figure \ref{SetUpPorAcase}. The employed water density refers to $1000$\;kg/m$^3$ and the dynamic viscosity is assigned to $\mu=0.001$\;Pa\;s. Pressure data was experimentally recorded for the steady state flow at four different locations to detect the free surface and is used for comparison in this study.

\subsubsection{A mesh resolution study}
\begin{figure}[htbp]
\centering
\includegraphics[scale=.4]{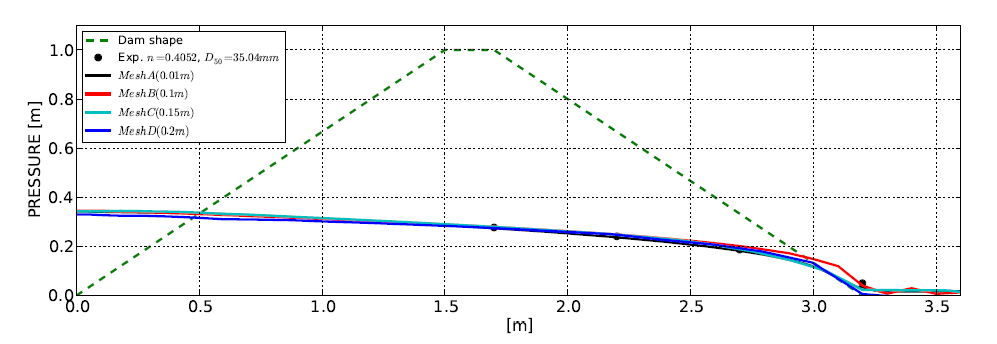}
\caption{Resolution study conducted with edge-based level set solver for the inclined porous dam case from Larese \cite{Larese2015}.}
\label{MeshStudyKratosAcaseOne}
\end{figure}

\begin{figure}[htbp]
\centering
\includegraphics[]{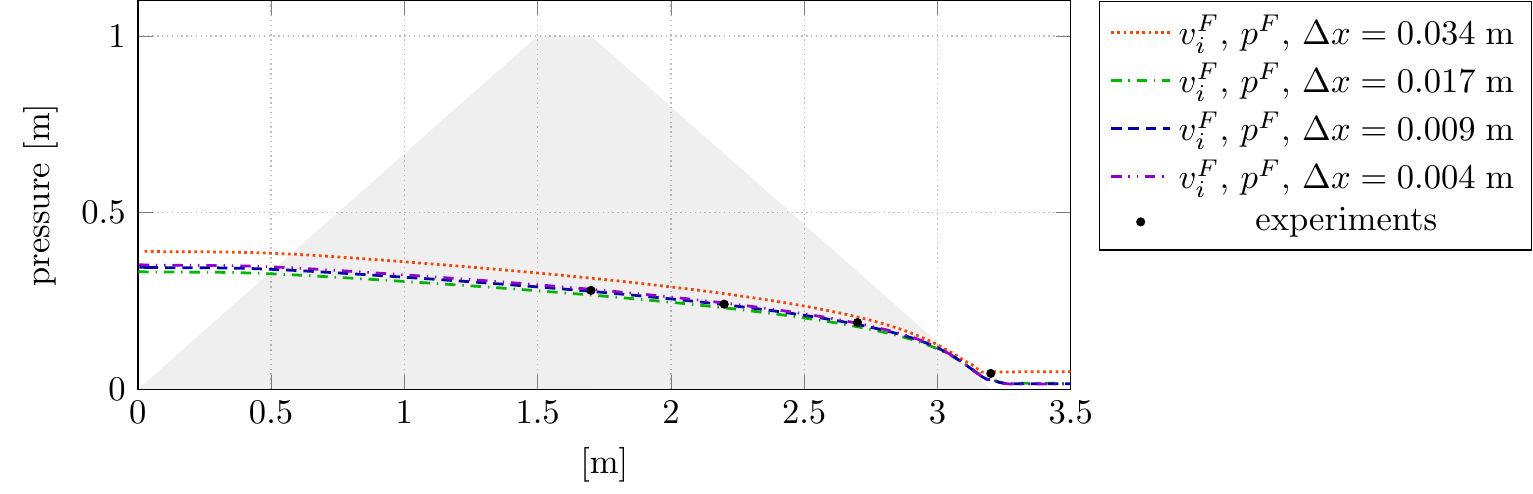}
\caption{Resolution study conducted with the $v_i^F$-based VoF formulation for the inclined porous dam case.}
\label{Acase_MeshStudyT200s_steadyInit}
\end{figure}

\begin{figure}[htbp]
\centering
\includegraphics[]{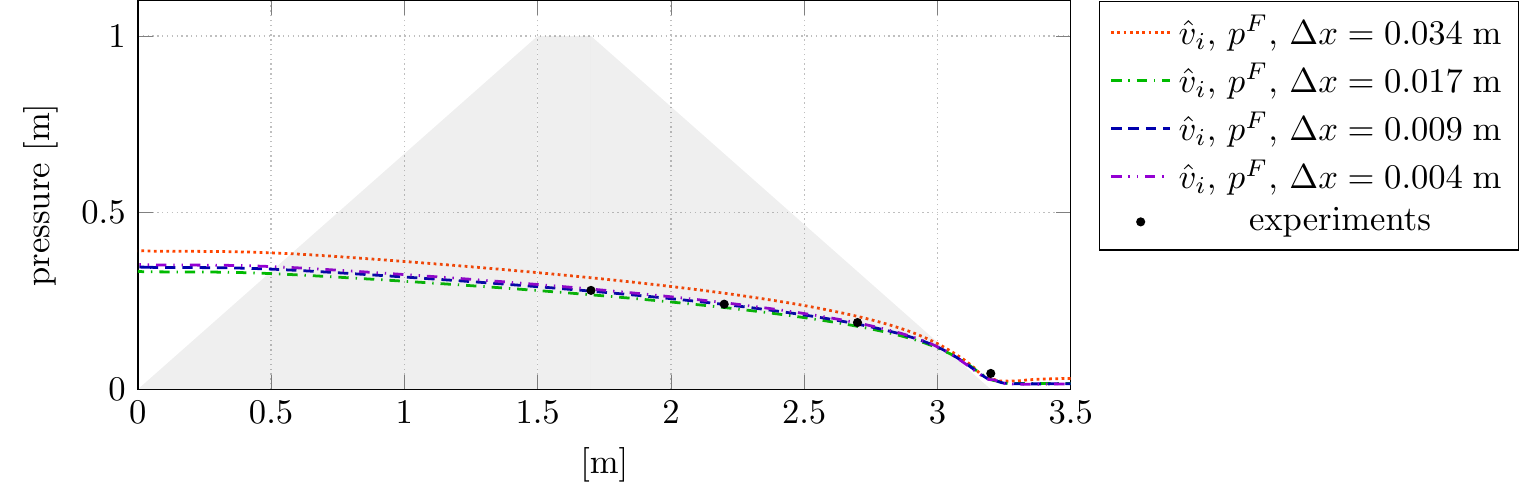}
\caption{Resolution study conducted with the $\hat v_i$-based VoF formulation for the inclined porous dam case.}
\label{Acase_MeshStudyT200s_steadyInit_VHatPF}
\end{figure}

was conducted for both investigated VoF formulations using homogeneous, isotropic Cartesian grids and compared with results of Larese et al. \cite{Larese2015}.
In comparison to the edge-based single-phase level set solver, cf. Fig. \ref{MeshStudyKratosAcaseOne}, 
both VoF formulations require significantly finer meshes to accurately resolve the liquid body (Figs. 
\ref{Acase_MeshStudyT200s_steadyInit}-\ref{Acase_MeshStudyT200s_steadyInit_VHatPF}). Mesh convergence is obtained with nodal distances of $\Delta x_i<1.0$\;cm and  $\Delta x_i<0.4$\;cm with the edge-based level set and  the VoF solver, respectively. The need for a better resolution  can be explained by the use of a true two-phase flow application. However, major predictive differences between the VoF results and the  edge-based level set benchmark are confined to grids with $\Delta x_i>1.7$\;cm.\\

\subsubsection{Numerical results}
\begin{figure}[htbp]
\centering
\includegraphics[]{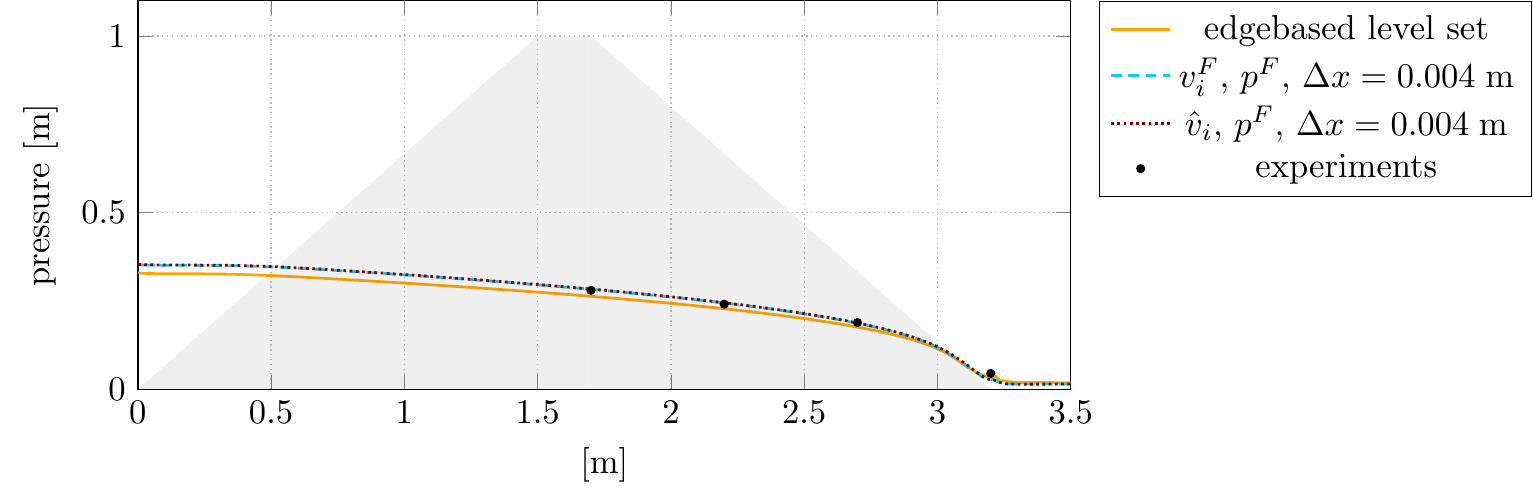}
\caption{Comparison of the predicted and measured free surface for the steady state of the flow through an inclined porous dam.}
\label{Acase_DiffSimulationsT200s}
\end{figure}

\begin{figure}[]
\begin{center}
\subfigure[$t=10$\;$\si{\second}$]{ 
\includegraphics[]{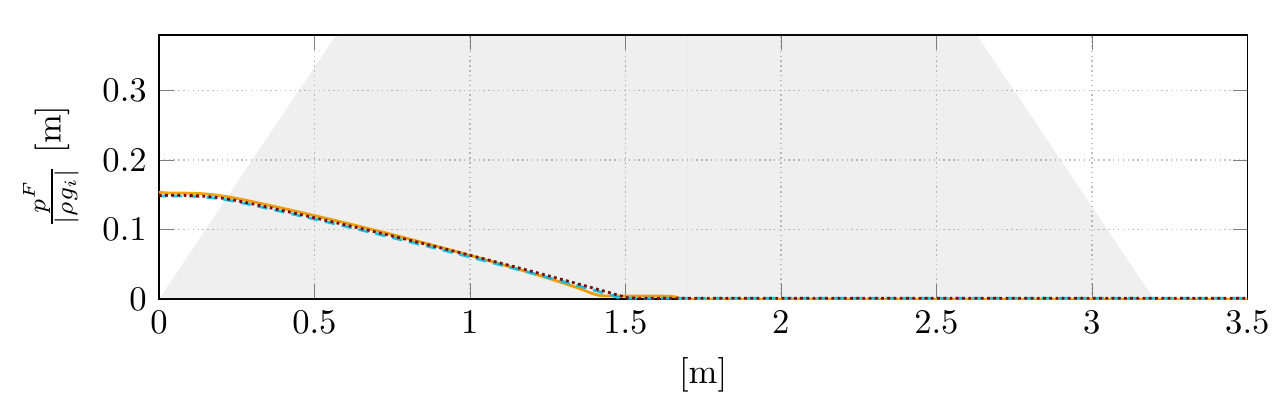}
}
\subfigure[$t=25$\;$\si{\second}$]{ 
\includegraphics[]{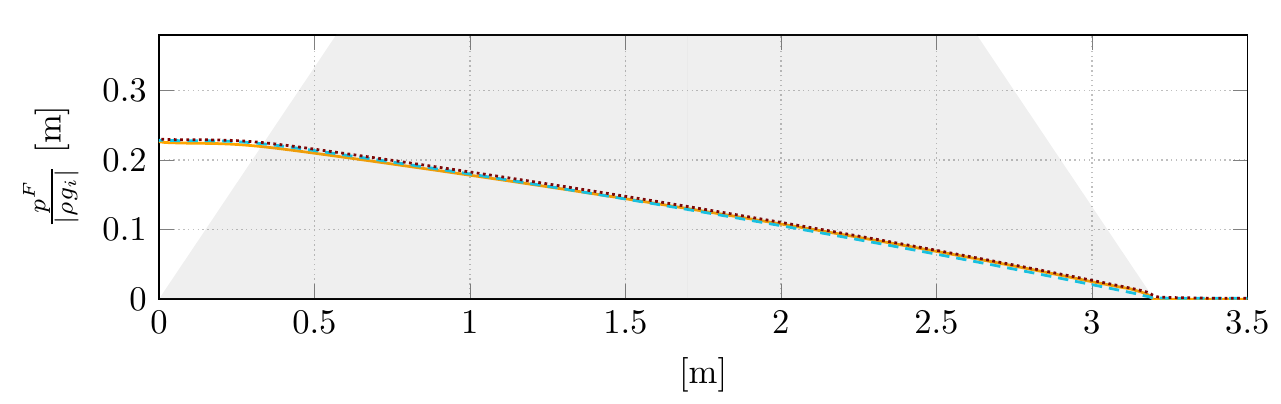}
}
\subfigure[$t=50$\;$\si{\second}$]{
\includegraphics[]{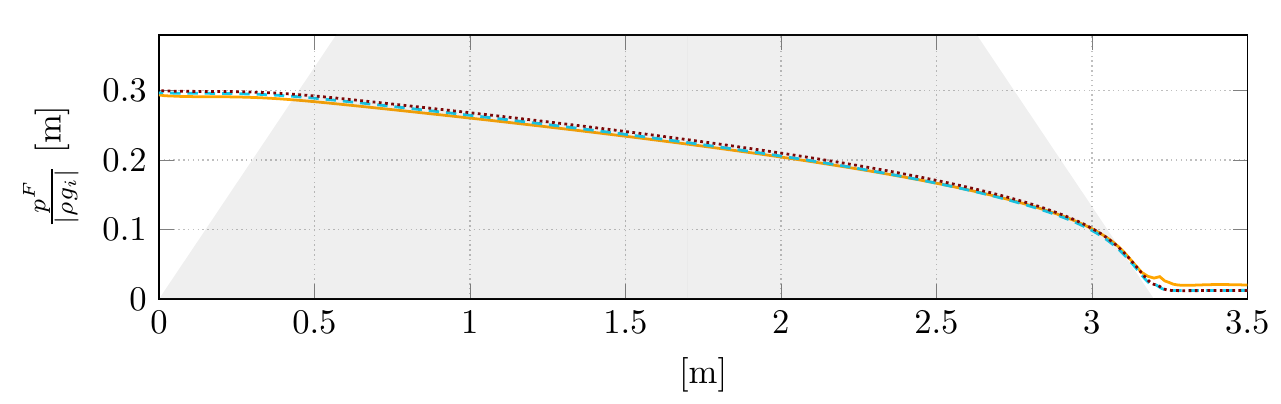}
}
\subfigure[$t=75$\;$\si{\second}$]{
\includegraphics[]{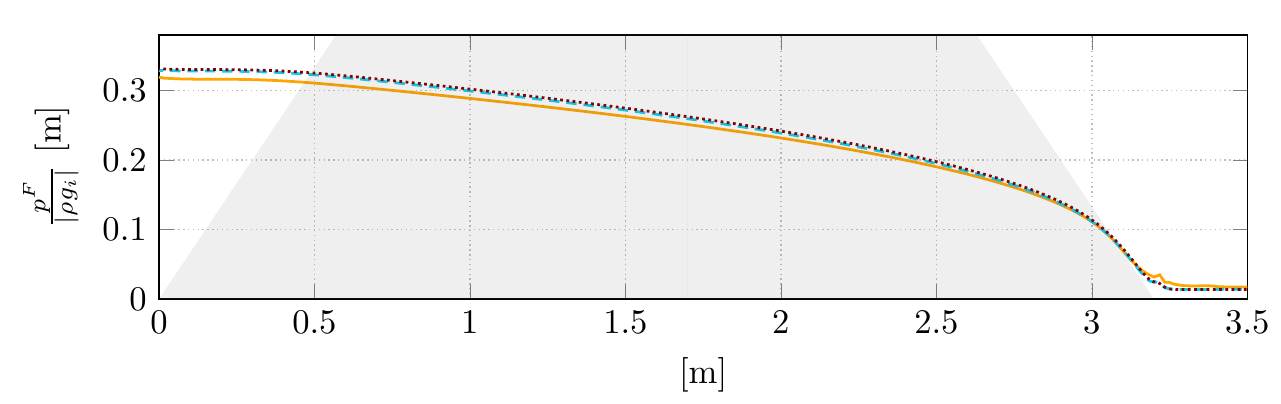}
}
\includegraphics[]{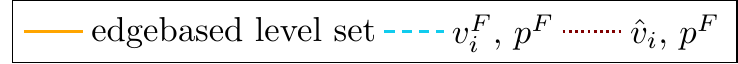}
\caption{Unsteady flow through an inclined porous dam. Comparison between $v^F_i$-based VoF results, 
$\hat{v}_i$-based VoF results 
and predictions of the edge-based level set solver 
.}
\label{Acase_Instat_DiffSimulations}
\end{center}
\end{figure}

are  compared with experimental data for the steady state in Fig. \ref{Acase_DiffSimulationsT200s}.
 Moreover, Fig. \ref{Acase_Instat_DiffSimulations} provides a comparison of the numerical data for the transient build up of the flow. Both aspects are discussed with regards to shape of the free surface, particularly in the porous regime. Experimental data is available at three locations inside the downstream half of the dam. A supplementary measurement position is located just downstream the dam. When attention is directed to the steady state depicted in Fig.  \ref{Acase_DiffSimulationsT200s}, the free surface shapes returned by the investigated two FV formulations display hardly any difference and agree very well with measured data. Therefore the formulation differences  are deemed insignificant for this case. The  edge-based solver predicts a free surface  which is located slightly below the VoF solutions and -- on average -- slightly underestimates the experimentally reported elevation. The steady state findings are confirmed by the comparison of transient results. Figure \ref{Acase_Instat_DiffSimulations} reveals that the free surfaces obtained from the FV and the edge-based solvers travel with a similar forward speed through the porous material, but a vertical/up lift tendency is observed by the VoF simulations when compared to the edge-based results. The latter might be attributed to the different formulation of the convective term and an increase of the porosity index in vertical direction.  

\subsection{3D Dam Break}
\label{3Dtestcases}
\begin{figure}[]
\begin{center}
\subfigure{
    \includegraphics[scale=.35]{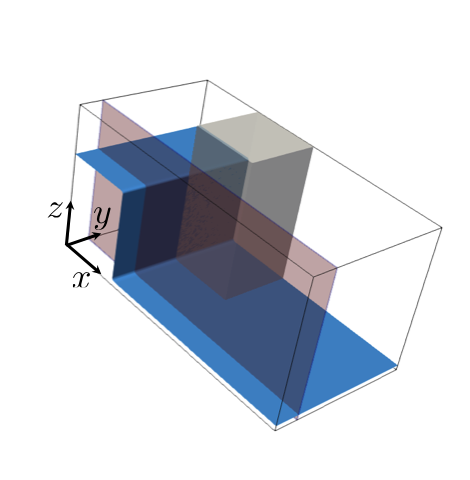}}
\subfigure{
\includegraphics[scale=.35]{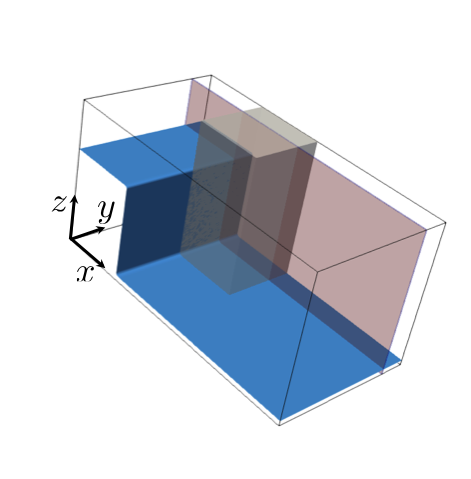}}
\subfigure[$y=10$\;$\si{\centi\metre}$]{ 
\includegraphics[]{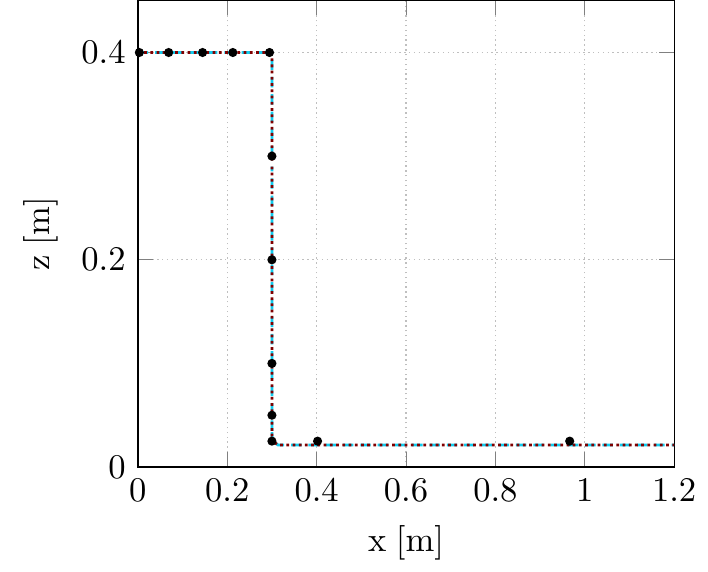}
}
\subfigure[$y=50$\;$\si{\centi\metre}$]{ 
\includegraphics[]{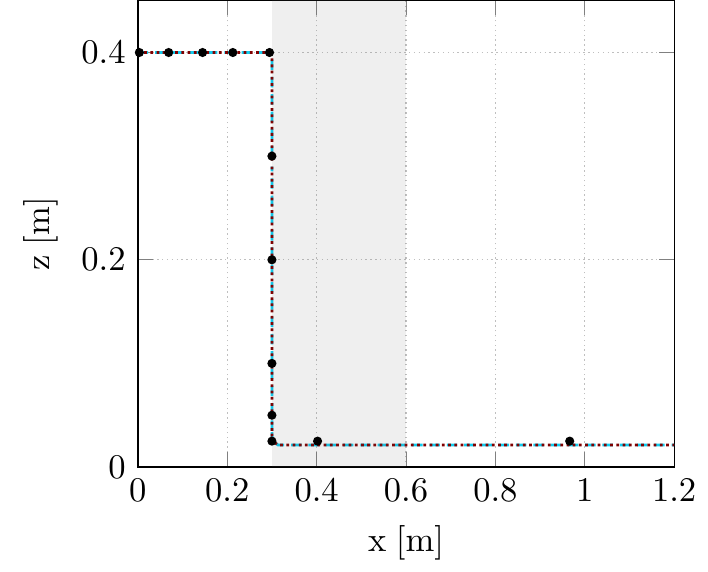}
}
\caption{Initial state of 3D example with a domain size $1.2$\;m$\times$$0.6$\;m$\times$$0.6$\;m. Porous dam expanding from $x=0.3$\;cm to $x=0.6$\;cm and $y=0.3$\;cm to $y=0.6$\;cm. Top: three dimensional views of initial state. Bottom: initial free surfaces in planes at $y=10$\;cm and $y=50$\;cm.}
\label{Init3D}
\end{center}
\end{figure}

To demonstrate the influence of different formulations for a more dynamic flow, a 3D dam break case reported by Del Jesus et al. \cite{DelJesus2012} is studied. 
The case is illustrated in Fig. \ref{Init3D} and refers to a three-dimensional extension of the crushed rocks case discussed in Sec. \ref{CrushedRocks}.  Additional to the HRIC scheme, an explicit interface sharpening algorithm described in \cite{Manzke2019} was employed for this case to diminish the interface smearing in later time steps. Results for both VoF formulations described in Sec. \ref{NUM} are compared against predictions of Del Jesus et al. \cite{DelJesus2012}. Unfortunately, no experimental data is available for this test case. 
A porous dam consisting of the crushed rocks material described in \ref{CrushedRocks} is positioned as given in Fig. \ref{Init3D} in a domain which is $0.6$\;m high, $0.6$\;m wide and $1.2$\;m long.
The dam features a height of $0.6$\;m, a width of $0.3$\;m and a length of $0.3$\;m. The initial water column is $0.6$\;m wide, $0.3$\;m long and $0.4$\;m high. As in the 2D cases, a free surface height of $2.5$\;cm is initiated in the rest of the domain. For the space/time-averaged Darcy velocity this yields  a pore Reynolds number $Re_P=1761$. Using the maximum Darcy velocity observed in space and time, the pore Reynolds number reads 
$Re_P=15099$. To obtain a similar resolution as \cite{DelJesus2012} the homogeneous horizontal grid spacing was assigned to $\Delta x=\Delta y=1$\;cm and the vertical to $\Delta z=0.5$\;cm.
\begin{figure}[]
\begin{center}
\subfigure[$t=0.08$\;$\si{\second}$, $y=10$\;$\si{\centi\metre}$]{ 
\includegraphics[]{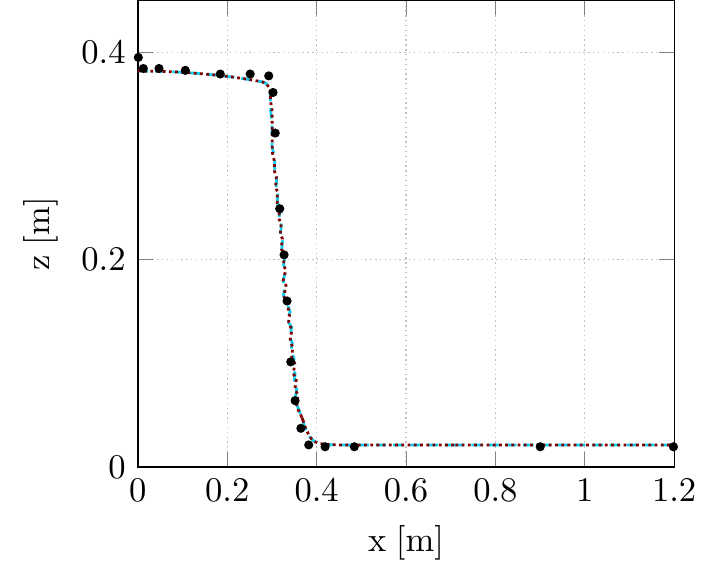}
}
\subfigure[$t=0.08$\;$\si{\second}$, $y=50$\;$\si{\centi\metre}$]{ 
\includegraphics[]{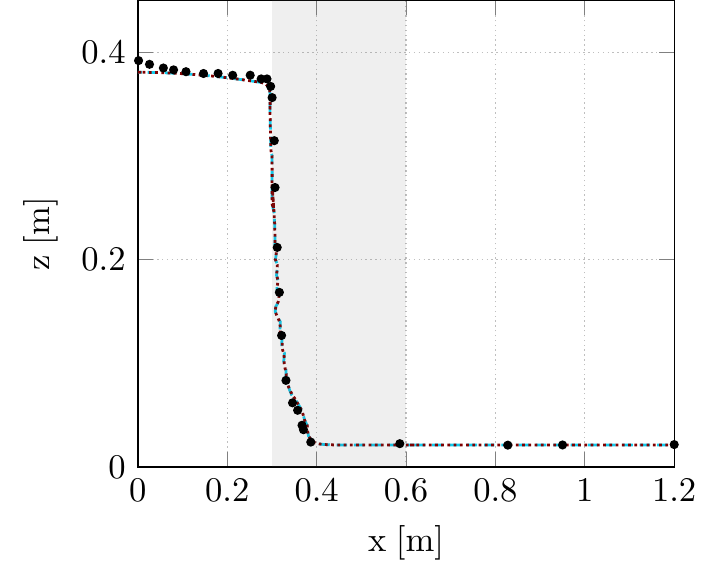}
}
\subfigure[$t=0.50$\;$\si{\second}$, $y=10$\;$\si{\centi\metre}$]{ 
\includegraphics[]{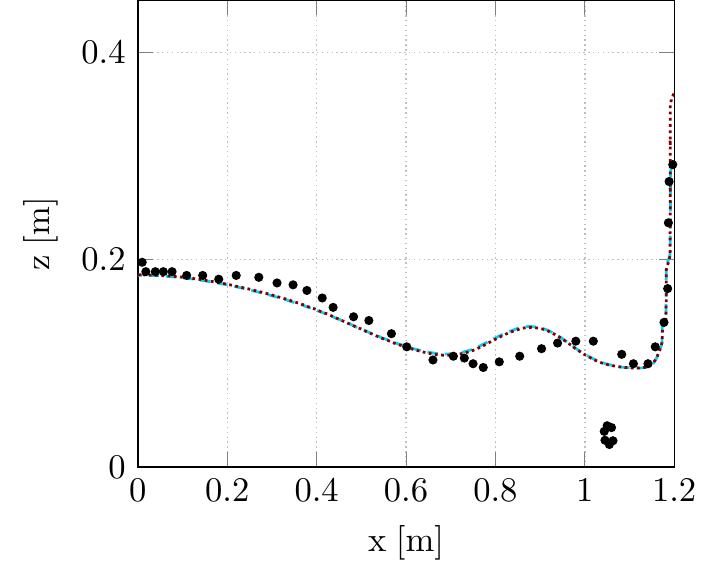}
}
\subfigure[$t=0.50$\;$\si{\second}$, $y=50$\;$\si{\centi\metre}$]{ 
\includegraphics[]{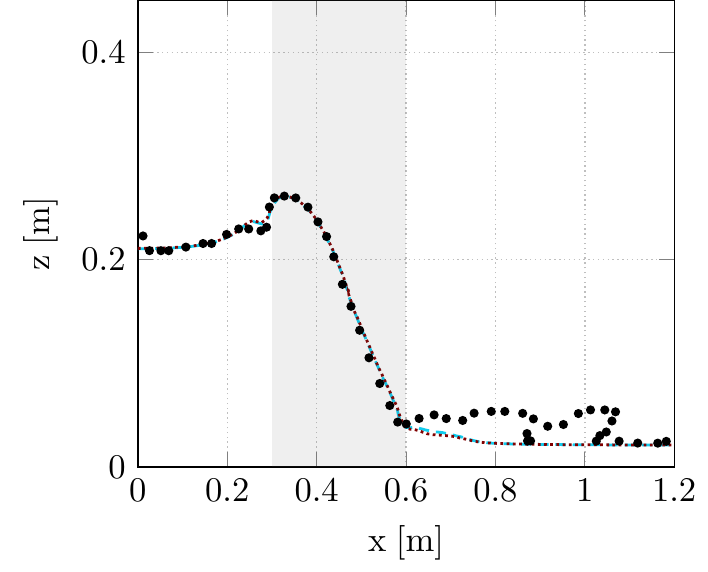}
}
\subfigure[$t=1.38$\;$\si{\second}$, $y=10$\;$\si{\centi\metre}$]{
\includegraphics[]{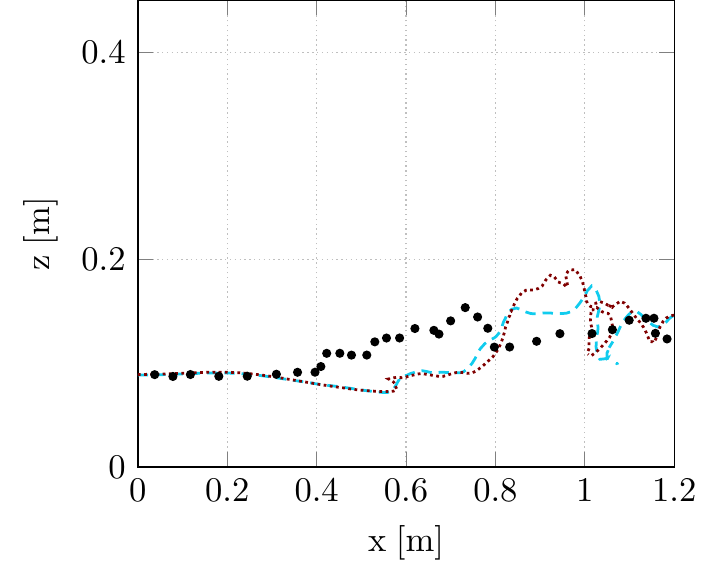}
}
\subfigure[$t=1.38$\;$\si{\second}$, $y=50$\;$\si{\centi\metre}$]{
\includegraphics[]{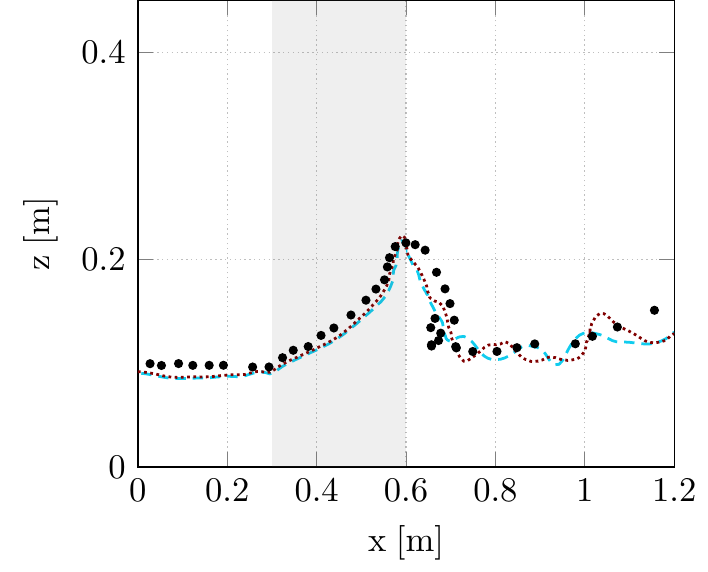}
}
\includegraphics[]{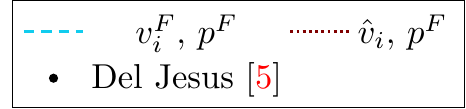}
\caption{3D porous dam break test case ($h=40$\;$\si{\centi\metre}$, $\Delta z=0.5$\;cm, HRIC + EIS) for different formulations.} 
\label{PorDamBreak3D_linInterp}
\end{center}
\end{figure}
The initial situation follows from Fig. \ref{Init3D}. A comparison of the two present VoF predictions with results reported in Del Jesus et. al \cite{DelJesus2012} is depicted in Fig. \ref{PorDamBreak3D_linInterp} for two lateral planes at $y=10$\;$\si{\centi\metre}$ and $y=50$\;$\si{\centi\metre}$. Both VoF methods discussed in Sec. \ref{NUM} deliver very close results. The difference in the formulation of the diffusion term in \eqref{XMomEq1a} and \eqref{XMomEq2a} seems to cause only minor deviations at the boundaries of the porous dam. \\
A reasonable overall agreement of the free surface development inside the porous material is found for all time steps and all displayed results.
The main deviations between the present results and \cite{DelJesus2012} are an upstream shift in the free surface elevation of the present predictions at $t=0.5$\;s in the plane which does not cut through the porous dam. This indicates an increased damming experienced in the present predictions, which also explains why more water is found at $t=0.5$\;s in the results of \cite{DelJesus2012} in the $y=50$ cm plane the dam. This indicates that a different mass transport through the porous boundaries is caused by the dissimilar treatment of the continuity equation. The same effect is thought to be responsible for the substantial differences of the observed free surface elevation at $t=1.38$\;s next to the dam.

\begin{figure}[htb]
\begin{center}
\subfigure[Del Jesus et. al \cite{DelJesus2012}]{
\includegraphics[scale=0.78]{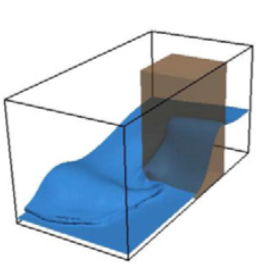}}
\subfigure[Larese et. al \cite{Larese2015}]{
\includegraphics[scale=0.5]{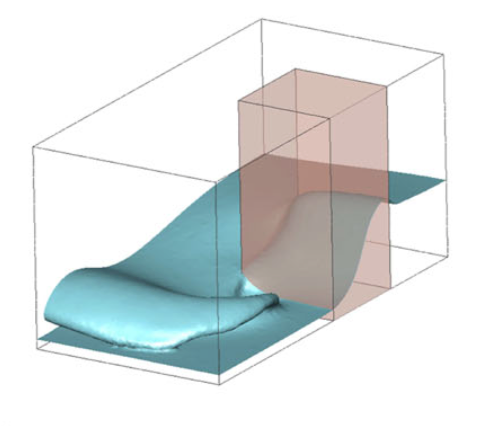}}
\subfigure[$\hat{v}_i$, $p^F$]{
\includegraphics[scale=.2]{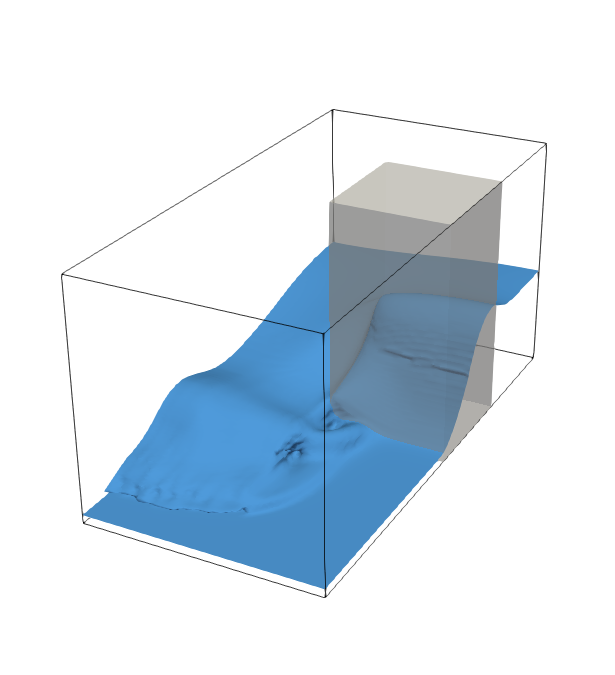}}
\subfigure[Del Jesus et. al \cite{DelJesus2012}]{
\includegraphics[scale=0.78]{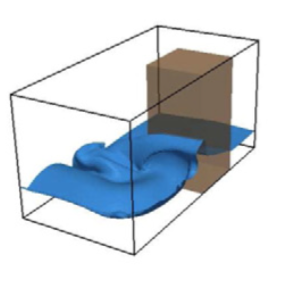}}
\subfigure[Larese et. al \cite{Larese2015}]{
\includegraphics[scale=0.51]{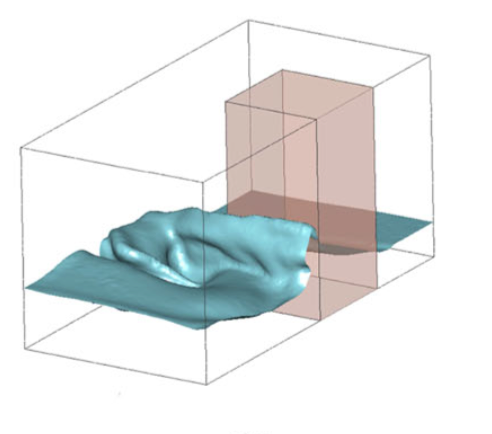}}
\subfigure[$\hat{v}_i$, $p^F$]{
\includegraphics[scale=.21]{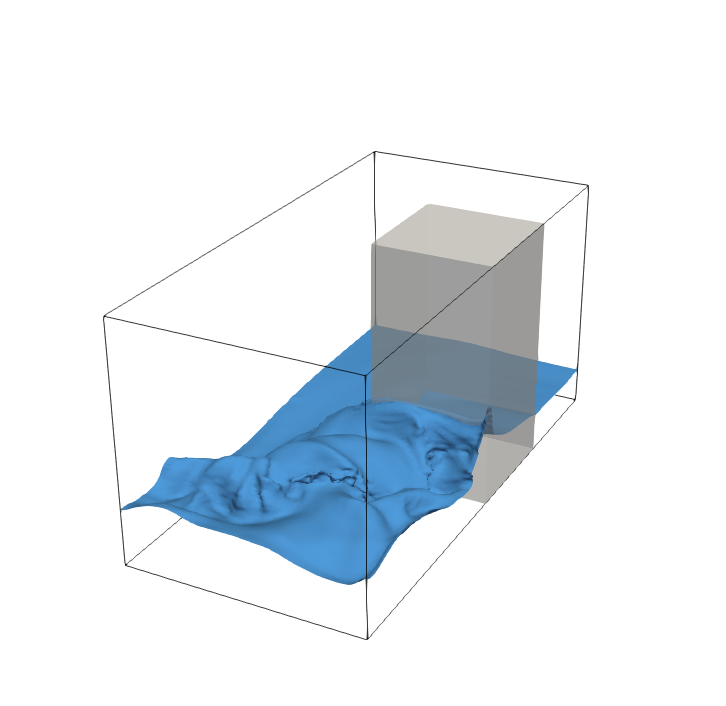}}
\caption{Comparison of free surfaces for time steps $t=0.40$\;s (top) and $t=1.33$\;s (bottom) for 3D porous dam test case.}
\end{center}
\end{figure}

\section{Conclusion}
\label{Concl}
The aim of this work was to assess different formulations of the governing momentum equations for free-surface flows  through rigid porous material. To this end, two exemplary formulations based on the Darcy and the fluid velocity were implemented in a VoF-based finite-volume procedure, and simulation results for 2D and 3D cases were compared against results from other literature reported formulations. It is found that for flows at porous Reynolds numbers up to $Re_p \le 5000$, formulation based changes of momentum transport terms  do only yield minor if not negligible deviations of the results, which are usually concealed by even subtle differences of the employed resistance law and the related parameters. Therefore, the particular choice of the employed momentum equations follows different motivations and further simplifications might be defensible from an engineering perspective. In conclusion, the optimal formulation might depend on the structure of the employed simulation procedure and the attainable efficiency and robustness. In this regard, an assessment of present run-times and robustness points to a preference of the Darcy velocity $\hat v_i$-based formulation.\\

\end{document}